\documentclass[preprint]{aastex631}

\usepackage{soul}

\usepackage{amsmath}

\begin{document}

\title{On the correlation between young massive star clusters and gamma-ray unassociated sources}

\author[0000-0003-3255-0077]{Giada Peron}
\affiliation{INAF Osservatorio Astrofisico di Arcetri Largo Enrico Fermi,5, 50125, Firenze, Italy}
\affiliation{Université Paris Cité, CNRS, Astroparticule et Cosmologie, 10 Rue Alice Domon et Léonie Duquet, F-75013 Paris, France}

\author[0000-0002-5014-4817]{Giovanni Morlino}
\affiliation{INAF Osservatorio Astrofisico di Arcetri Largo Enrico Fermi,5, 50125, Firenze, Italy}

\author{Stefano Gabici}
\affiliation{Université Paris Cité, CNRS, Astroparticule et Cosmologie, 10 Rue Alice Domon et Léonie Duquet, F-75013 Paris, France}


\author[0000-0002-9881-8112]{Elena Amato}
\affiliation{INAF Osservatorio Astrofisico di Arcetri Largo Enrico Fermi,5, 50125, Firenze, Italy}

\author{Archana Purushothaman}
\affiliation{Dipartimento di Fisica e Astronomia "Augusto Righi", Universit\`{a} di Bologna, Via Piero Gobetti, 93/2, 40129 Bologna, Italy}

\author{Marcella Brusa}
\affiliation{Dipartimento di Fisica e Astronomia "Augusto Righi", Universit\`{a} di Bologna, Via Piero Gobetti, 93/2, 40129 Bologna, Italy}\affiliation{INAF Osservatorio di Astrofisica e Scienza dello Spazio Bologna, Via Piero Gobetti, 93/3, 40129 Bologna, Italy}

\begin{abstract}
Star clusters (SCs) are potential cosmic-ray (CR) accelerators and therefore are expected to emit high-energy radiation. However, a clear detection of gamma-ray emission from this source class has only been possible for a handful of cases. This could in principle result from two different reasons: either detectable SCs are limited to a small fraction of the total number of Galactic SCs, or gamma-ray-emitting SCs are not recognized as such and therefore are listed in the ensemble of unidentified sources. In this Letter we investigate this latter scenario, by comparing available catalogs of SCs and H\textsc{ii} regions, obtained from Gaia and WISE observations, to the gamma-ray GeV and TeV catalogs built from {\it Fermi}-LAT, H.E.S.S. and LHAASO data. The significance of the correlation between catalogs is evaluated by comparing the results with simulations of synthetic populations. A strong correlation emerges between {\it Fermi}-LAT unidentified sources and H\textsc{ii} regions which trace massive SCs in the earliest ($\lesssim 1-2$ Myr) phase of their life, where no supernova explosions have happened yet, confirming that winds of massive stars can alone accelerate particles and produce gamma-ray emission at least up to GeV energies.
The association with TeV-energies sources is less evident. Similarly, no significant association is found between Gaia SCs and GeV nor TeV sources. We ascribe this fact to the larger extension of these objects, but also to an intrinsic bias in the Gaia selection towards SCs surrounded by a lower target gas density, that would otherwise hinder the detection in the optical waveband.

\end{abstract}

\keywords{gamma rays: general; gamma rays: stars; (ISM:) cosmic rays; (ISM:) HII regions; (Galaxy:) open clusters and associations: general.}

\section{Introduction} \label{sec:intro}

Massive star clusters (SC) have been proposed as prominent cosmic-ray (CR) accelerators, possibly able to account for both the observed spectrum of these particles up to the knee \citep{Cesarsky1983a,Montmerle2010,Bykov2020,Vieu2022CosmicSuperbubbles, Vieu2023MassiveEnergies}, and for some anomalies in their composition \citep{Tatischeff2021TheComposition,Gabici2023SCs}.
The number of SCs detected in gamma rays however remains small (about a dozen) compared to the number of SCs that populate the Milky Way. At the same time, more than half of the sources in gamma-ray catalogs from MeV to PeV energies remain unidentified \citep{Abdollahi2022IncrementalCatalog,Abdalla2018,Cao2023Catalog}.  SCs might be particularly elusive as gamma-ray emitters because of their intrinsically low surface brightness, due to their large extension, or as a result of source confusion caused by neighboring emitters, that might partly or completely obscure the SCs' emission. But, it is also possible that gamma-ray emitting SCs are not identified as such because a systematic association with this source class was never attempted and therefore {many of them are probably hidden under the flag of \textquoteleft unassociated' sources. }
This was the case for example for the clusters unveiled in the Vela region \citep{Peron2024ThePopulation} thanks to their spatial coincidence with unassociated {\it Fermi}-LAT sources. In that case, a careful morphological analysis was needed to clarify the nature of the emitter. However, a statistical comparison among catalogs of gamma-ray sources and catalogs of SCs can readily provide hints on whether the latter can be considered as gamma-ray emitters and it is the aim of this work.

For these reasons, we consider here two catalogs of SCs: the first based on observations of the WISE infrared satellite at 22~$\mu$m \citep{Anderson2014TheRegions}, unveiling 8412 Galactic H\textsc{ii} regions, and the second one based on Gaia data, tracing star positions, ages and velocities \citep{Cantat-Gaudin2020}, in the optical band.

\section{Associations}
In order to evaluate the associations and the degree of chance-coincidence we proceed as follows. First, we match sources between different catalogs based on positional coincidence within a certain radius, $R_{*}$, whose definition depends on the considered catalog, and reflects the presumed origin of the emission for systems in different evolutionary stages.

Embedded SCs are enclosed with dense material that provides target for the interaction, and therefore the emission is expected to be rather compact. 
For this reason, for the clusters of the WISE catalog we assume the search radius to be coincident with the extension of the H\textsc{ii} region measured in the infrared $R_{*}=R_{\rm HII}$. Older SCs, instead, are presumably surrounded by the low-density bubble blown by their winds.
 
\begin{figure}
    \centering
    \includegraphics[width=0.7\linewidth]{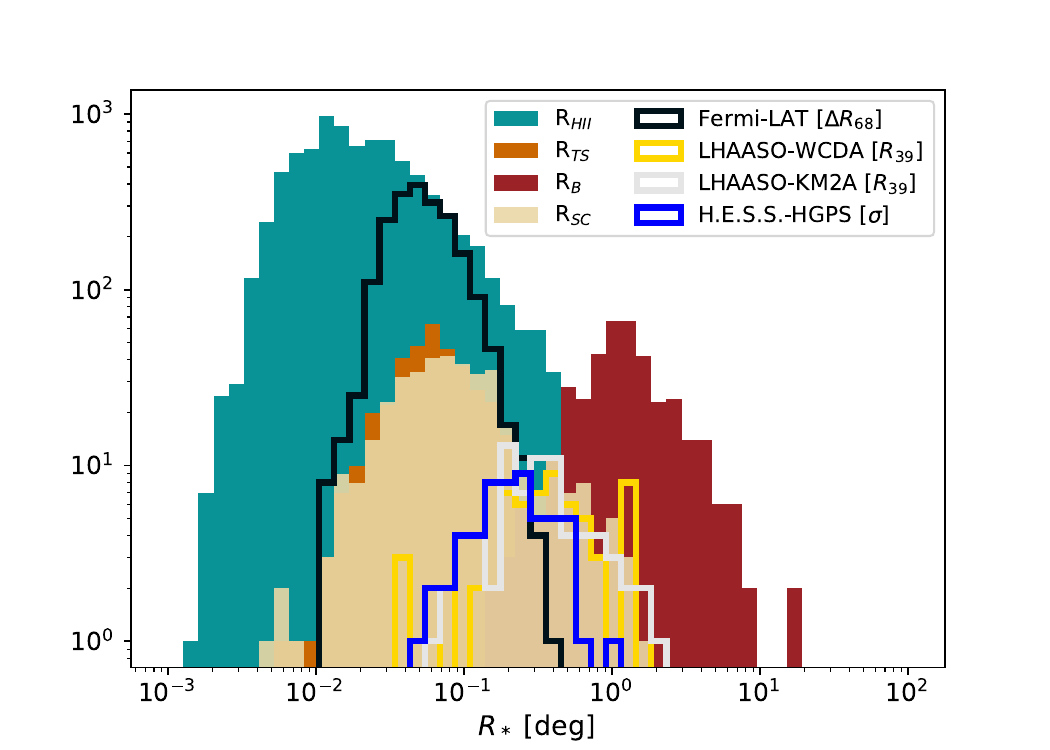}
    \caption{Histogram of sizes compared in this work.  {For the SCs, we report the size of the H\textsc{ii} regions, $R_{HII}$, the SC radius, $R_{\rm SC}$, defined as the radius that contains 50\% of member stars, and $R_{\rm TS}$ and $R_{B}$, defined in the text. For the gamma-ray sources we report different sizes: being {\it Fermi}-LAT unassociated sources mostly point-like, we report their positional uncertainty, derived from the 68\% confidence ellipse; LHAASO and H.E.S.S. sources are instead fitted with Gaussians and the $\sigma$ and the 39\% containment radius are reported.  }}
    \label{fig:sizes}
\end{figure}

According to Weaver's theory of wind-blown bubbles \citep{Weaver1977InterstellarEvolution.} the size of the bubble, $R_{\rm B}$, scales with the SC age, $t_{\rm SC}$: $R_{\rm B} \propto (L_{\rm w} t_{\rm SC}^{3} n_0^{-1})^{1/5} $, where $L_{\rm w}=0.5 \dot{M} v_{\rm w}^2$ is the mechanical power of the SC wind,  {that depends on the mass-loss rate, $\dot{M}$, and on the wind velocity, $v_w$. $n_0$ is the external density of the medium, that we assume to be 10~cm$^{-3}$, compatible with average values found for molecular clouds in the Galaxy \citep{Miville-Deschenes2016}}. 
{Inside the bubble, the pre-existing gas has been swept towards the outer edge and only shocked wind material is expected to be present, 
resulting in a rather low density $\lesssim~0.1$~cm$^{-3}$, that suppresses the chances for interactions, unless the target is increased by evaporation of the gas of the cold shell or by the presence of clumps \citep{Menchiari2023}.  }
In this scenario, then, accelerated particles propagate from the termination shock, whose location also depends on the properties of the SC ($R_{\rm TS} \propto\dot{M}^{3/10} v_{\rm w}^{1/10} t_{\rm SC}^{2/5} n_0^{-3/10} $) {to distances where they find} target for interaction: a dense gas region, or a high radiation field. While dense gas regions are mostly expected at the bubble boundary, large radiation fields are expected near the stars {(close to $R_{\rm SC}$, the radius of the stellar core)} and therefore the exact location of the gamma-ray emission will depend on the dominant emission mechanism. 
To search for gamma-ray counterparts of similar objects we therefore looked for overlapping sources within these different radii  {($R_*=R_{\rm TS}, R_{\rm B},R_{\rm SC}$)}. Recently, \cite{Celli2023} provided an estimate of the mass and wind properties for 387 of the youngest ($<$ 30 Myr) of the Gaia-identified SCs, {that allows us to estimate $R_B$ and $R_{TS}$  (see Figure \ref{fig:sizes}).  The age selection corresponds to} the time interval during which a cluster is expected to emit in gamma rays  either due to winds or to SNe, being 30 Myr the age at which 8 M$_\odot$-stars, the least massive SN progenitors, explode. After that, both the power in winds and in supernova remnants significantly declines \citep{Vieu2022CosmicSuperbubbles}. {Therefore, gamma-ray emission emerging from older SCs is most likely associated to different objects, such as millisecond pulsars \citep{GlobularSCs}}.  

\subsection{\textit{Fermi}-LAT}\label{sec:gev}
The \textit{Fermi}-LAT latest data release \cite[DR3;][]{Abdollahi2022IncrementalCatalog}\footnote{A newer version of the data catalog has been released in the form of a pre-print. We repeated the same procedure with this updated catalog and verified that the conclusion is unchanged.}
contains 2082 unassociated sources, where sources are here considered unassociated if they don't have any reported counterpart at any confidence level\footnote{The association in the catalog is based on spatial arguments and its significance is evaluated with a Bayesian: associations with the highest-confidence are indicated as ‘CLASS~1', while lower-confidence level associations are also reported and flagged as ‘CLASS~2'. 
{The catalog counts 2157 unassociated sources, but among these, 209 sources have a low-confidence association, therefore we exclude them from this study;} instead we include the 134 sources that are flagged as unknown {in the 'CLASS~1' or 'CLASS~2' entry}, namely those low-latitude ($|b|<10^\circ$) sources without a clear counterpart, but which are close to radio and X-ray sources. We refer to \cite{Abdollahi2022IncrementalCatalog} and references therein for further details }.
We match then the unassociated sources by demanding that the source center is located at a distance from the SC center smaller than $R_*$. Most of the considered Fermi-LAT sources are point-like, with the exception of 18 sources. 
For the matching procedure, we consider these too as point like and coincident with their centroid coordinates.\footnote{A posteriori we further checked for SCs overlapping these sources within an area corresponding to the extension measured by {\it Fermi}-LAT. In most cases no additional associations are found. In the few cases for which an overlap is found, this is with more than one SC. 
One instance is the 2.5-deg wide region 4FGL~J1036.3-5833e that hosts 70 HII regions and 13 Gaia SCs, including Westerlund~2. Another is the 1.3-deg wide 4FGL~J1109.4-6115e, overlapping 38 HII regions and 6 SCs including NGC~3603. These could indicate a diffuse emission around SCs or regions of enhanced star formation similar to the mini-starburst W43 \citep{Yang2020}, anyway complicating the association with the correct objects, so we do not account for them in this work. }

We show in Figure \ref{fig:sizes}  the positional uncertainty of the {\it Fermi}-LAT source positions, $\Delta R_{68}$, which is computed from the 68\% confidence ellipse, by considering a circle with the same area. We see that $\Delta R_{68}$ is in general smaller than, or comparable to, the average radius of the Gaia SCs, while it is on average larger than the WISE radius, meaning that the comparison with the latter is largely conservative. {Even if $\Delta R_{68}$ is not directly used in the matching procedure, we report its distribution to show that} Fermi-LAT sources could be displaced from WISE regions {even} only as a result of their poor localization.

\begin{figure}
    \centering
\includegraphics[width=0.58 \linewidth]{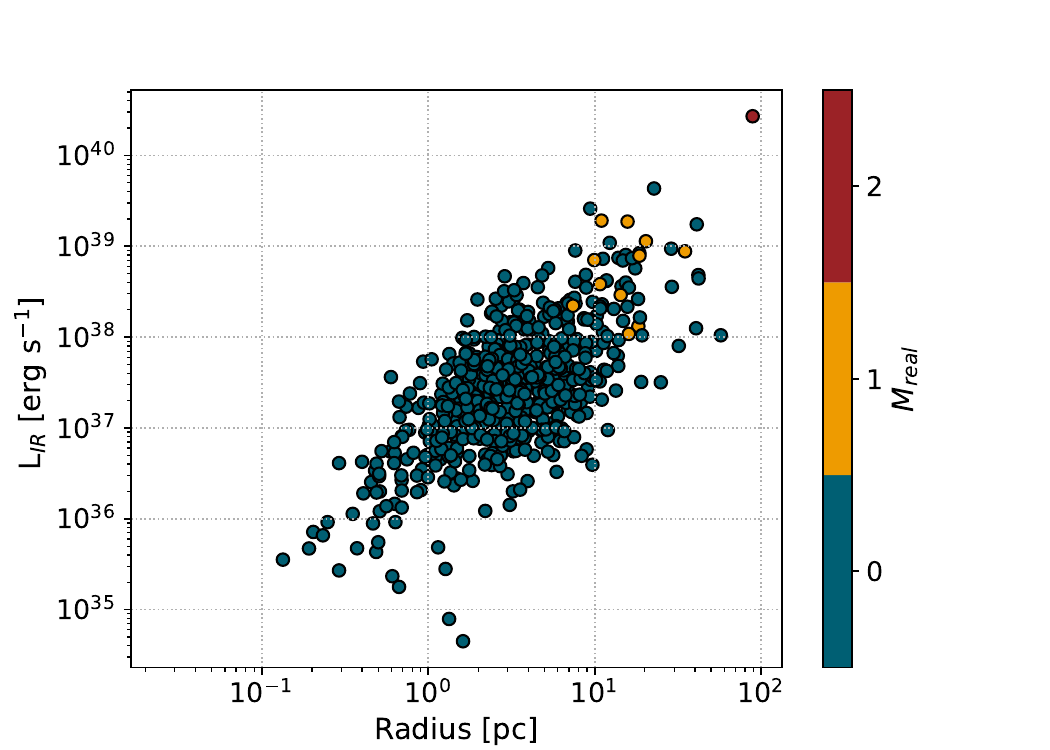}\includegraphics[width=0.41 \linewidth]{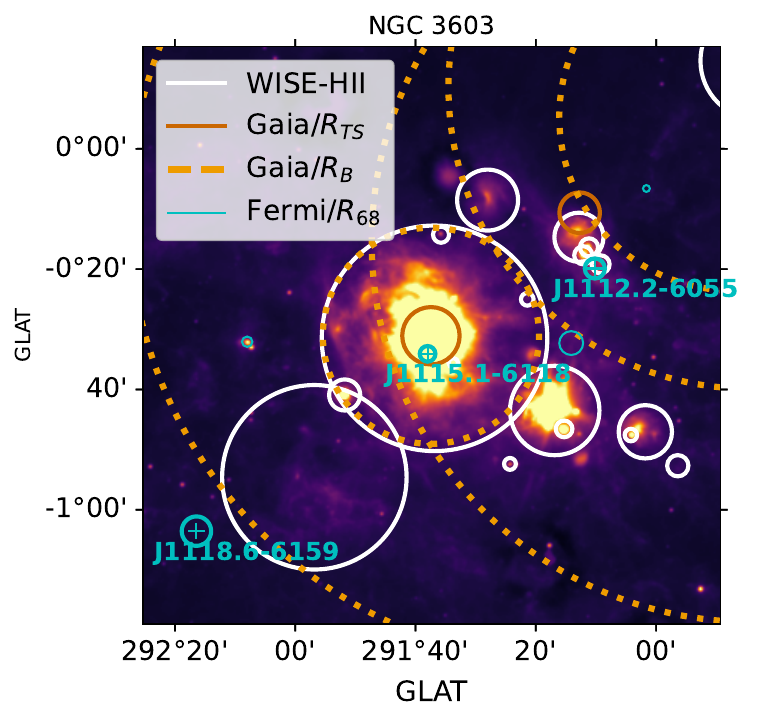}
    \caption{Left: Infrared (22$\mu$m) luminosity and extension of WISE HII regions for which we know the flux and distance. The color-code indicates the number of \textit{Fermi}-LAT unidentified sources that overlap these objects. Right: The WISE 22-$\mu$m map towards the young massive cluster NGC~3603. The HII regions considered in this work are drawn in white, while the Gaia SCs are drawn in orange, dashed orange lines are for $R_{B}$, solid orange lines are for  $R_{TS}$. Light blue circles indicate the Fermi 4FGL centroid uncertainty, $\Delta R_{68}$; among these, sources overlaid with a cross are unassociated. For the latter, the name, as reported in the catalog, is also reported } 
    \label{fig:Lir_radius}
\end{figure}

Still, we find 127 WISE objects ($\sim$ 1.5 \% of 8412 listed in the catalog) that have at least one overlapping \textit{Fermi}-LAT source; in a dozen cases more than one unidentified source overlaps a single H\textsc{ii} region. 
Overall we find that a total of 138 Fermi unassociated sources ($\sim$7 \% of the total) are potentially associated to young embedded stellar clusters. 
A morphological investigation case by case is needed to clarify whether the coincidence with multiple sources is due to an extended gamma-ray emission. As the regions increase in size, an increase of overlapping sources is expected for two reasons: a purely geometrical reason, because of the larger projected area (and this is tested in the evaluation of the significance of the correlation), and one physical reason as we expect that the most powerful SCs heat a larger surrounding region \citep{Stahler2004}. 
Unfortunately we do not have any information on the luminosity of the SCs embedded in our H\textsc{ii} regions, but for a fraction of those ($\sim$~1/8 of the sample) we have an estimated infrared luminosity in different bands \citep{Makai2017TheRegions} that we can use as a tracer of the wind power, since we expect that all emission is powered by, and therefore proportional to, the stellar luminosity. For a fraction of this sample, we also know the distance, so that we can estimate the intrinsic infrared luminosity, $L_{\rm IR}$, and the physical size. 
We consider here the emission at 22~$\mu$m, because it has been demonstrated that this is equivalent to other wavebands for tracing H\textsc{ii} regions around massive stars \citep{Makai2017TheRegions}.  Plotting the derived luminosity as a function of the extension of the considered regions (Fig. \ref{fig:Lir_radius}), one can see that the two quantities correlate, and that the number of overlapping \textit{Fermi}-LAT sources is higher for high-luminosity H\textsc{ii} regions, in agreement with the expectations.

In the case of Gaia SCs, we tested the associations within $R_{\rm SC}$, defined as the radius that contains 50\% of cluster members, and within $R_{\rm TS}$ and $R_{\rm B}$ defined before. 
{The right panel of Figure~\ref{fig:Lir_radius}, shows an example of {\it Fermi}-LAT sources matching with SCs in the region of the known young massive star cluster NGC~3603 \citep{yang2016NGC, Saha2020Morphological3603} We see that the the 4FGL source is contained at the center of the infrared emission\footnote{The background map has been obtained from \href{https://alasky.cds.unistra.fr/hips-image-services/hips2fits}{https://alasky.cds.unistra.fr/hips-image-services/hips2fits} a service provided by CDS.}, and within our estimated $R_{TS}$ and $R_{B}$}. {As we can see (also from Fig. \ref{fig:sizes}),} the latter is on average larger than the \textit{Fermi}-LAT positional errors. This can cause spurious associations that can be evaluated only by comparison with simulations, as discussed later. Indeed the number of clusters matching {\it Fermi}-LAT-unidentified sources is much larger when using $R_B$ rather than the other radii leading to an unreasonable fraction of $\sim$30\% of Fermi unassociated sources overlapping SCs. The numbers for the matches are reported in Table \ref{tab:nSCs}, in the Appendix. Instead, only a small fraction $\sim 0.4$\% of the GeV sample is found within SCs' termination shocks or the clusters' stellar cores. (see Appendix \ref{sec:nmatch} for further details).

In order to  assess the degree of chance coincidence of the derived matches, we compare the number of matches obtained with the real catalog, $M_{\rm real}$, and the number of matches obtained with simulated catalogs, $M_{\rm sim}$. We generated 1000 synthetic catalogs, constructed by performing a Monte-Carlo extraction over the longitude and latitude distributions of the \textit{Fermi}-LAT unidentified sources (see Appendix \ref{sec:montecarlo}). We keep fixed instead the position and the extension of the clusters from the catalogs and repeat the matching procedure with the same assumptions on $R_{*}$. We assess then the significance of the correlation as: 
\begin{equation}
  \Sigma=\frac{M_{\rm real} - \langle M_{\rm sim} \rangle }{\sigma_{\rm sim}}    
\end{equation}
where $\langle M_{\rm sim} \rangle$ and $\sigma_{\rm sim}$ are the mean and the standard deviation of the distribution of  $M_{\rm sim}$. {We define a good correlation between catalogs if $\Sigma>3$, that approximately corresponds to a probability of chance coincidence smaller than 1\% (see Appendix \ref{sec:montecarlo}) .}

\begin{figure}
    \centering
    \includegraphics[width=0.5 \linewidth]{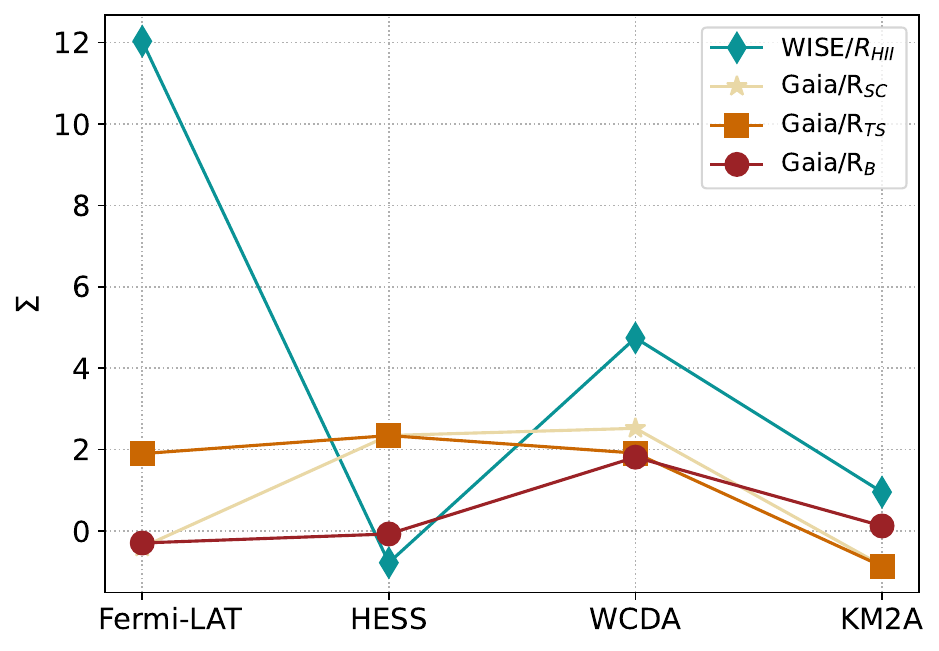}\includegraphics[width=0.5 \linewidth]{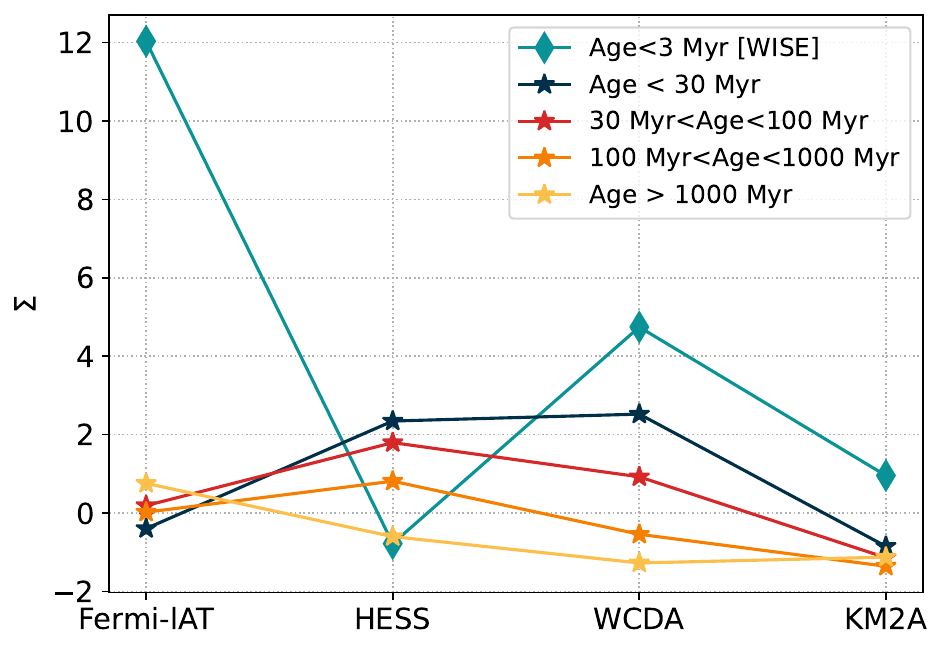}
    \caption{{Left:} significance of the correlation between the considered  star-clusters catalogs   and the different gamma-ray catalogs. The different colors refer to the different star-cluster samples and the different matching radii used in the matching, as indicated in the legend. {Right: significance of the correlation between gamma-ray source catalogs, and Gaia identified clusters of different ages. In this case, the radius used for correlation is the stellar core radius, $R_{SC}$.} }
    \label{fig:sig_all}
\end{figure}

The resulting significance of the correlation between the \textit{Fermi}-LAT catalog and the SC catalogs is reported in Figure~\ref{fig:sig_all}. {There, we also report on the significance of associations with different Gaia sub-samples, based on the SC's age and compare it with the significance obtained with WISE catalog, that should trace very young ($<$3 Myr) SCs. Since we do not have information about the luminosity of Gaia SC's beyond 30 Myr, we can use only $R_{SC}$ as matching radius. We see that only with the youngest SCs from Gaia ($<$ 30 Myr) or WISE, we obtain a considerable significance, that seems to decrease instead with the age of the cluster. For the $<$30 Myr sample, we further investigate the significance in smaller sub-samples based on age, and on wind luminosity (Fig. \ref{fig:sig_TeV}) }
Having the mass and the age of the clusters, we can also evaluate the probability that a supernova (SN) exploded in the cluster, as explained more in detail in Appendix \ref{sec:supernovae}, and test whether these events could enhance the correlation. 

For WISE SCs we {make some sub-samples using}, when available, the infrared luminosity of the regions evaluated from the reported flux and distances. {Moreover,} as a check for the possible biases in the choice of the matching radius, $R_*$, we evaluated the significance of the association by restricting to sub-samples based on their angular extensions. {Indeed, in case of random coincidence the significance would decrease with increasing radii, for geometrical reasons.} The results are also shown in Figure~\ref{fig:sig_TeV}. 

We see that the significance of association of WISE SCs is $\Sigma \sim 12$ for the whole catalog and always $\gtrsim 4$  for each investigated sub-sample. The significance increases with increasing $R_{\rm HII}$, clearing up doubts on a possible bias due to the larger matching radius. Even when the radius is smaller than the average \textit{Fermi}-LAT positional error, {$\sim 0.06 ^\circ$}, {and hence the matching could fail due to a poor localization}, the significance is higher than 4, suggesting a very high degree of connection between these source classes. {Even when halving the matching radius, the correlation is still significant ($\Sigma\sim 6$)}. Alongside, the significance increases with increasing  $L_{\rm IR}$, as expected for sources powered by stars.  

The association significance found for Gaia SCs is instead much lower, and always $\lesssim 3$, regardless of the chosen matching radius. The highest significance is found when matching with the termination shock radius, however the results don't allow us to draw a firm conclusion on this point. A slight increase of the significance can be seen as a function of the wind luminosity, but again the numbers don't allow us to establish a  correlation between the two populations.

\subsection{Higher energies: H.E.S.S.,and LHAASO}\label{sec:tev}
At higher energies, the most comprehensive catalogs of sources have been obtained with the H.E.S.S., HAWC and LHAASO experiments. The first collected, in the HGPS \citep{Abdalla2018}, the sources observed within $60^\circ<l<250^\circ$ and $|b|<3^\circ$, while LHAASO and HAWC surveyed the northern sky approximately at all accessible longitudes and latitudes.
For this work we considered the HGPS catalog and the 1st LHAASO catalog \citep{Cao2023Catalog}. The identification of these TeV sources is still uncertain: more than half (48/78) of the HGPS sample is composed of unidentified sources, that we use here for the catalog matching. The LHAASO catalog contains 68 sources revealed with the WCDA detector in the 1-10 TeV energy range and 75 sources revealed with KM2A, the ultra-high energy part of the observatory that goes beyond 100 TeV. Remarkably, 44 of these sources are revealed above this energy and are then classified as ultra-high energy emitters.  Among the LHAASO emitters, only the Crab nebula is firmly associated, while the nature of the other sources remains unconstrained. A few studies \citep{Cao2023Catalog,Emma2022} have attempted  an association with pulsars to try to explain the extended very-high-energy emission in the context of pulsar wind nebulae, and have found a high level of coincidence with strong pulsars. Nevertheless the association with SCs cannot be excluded \textit{a priori} and we try to quantify it here.

Differently from {\it Fermi} sources, most TeV sources are extended, with an extension that is larger on average than the size of the considered WISE H\textsc{ii} regions and comparable to that of the Gaia SCs, as can be seen in Fig. \ref{fig:sizes}. For this reason we cannot consider the sources to be point-like when evaluating the association, and we proceeded by including the extension when matching both with the real and simulated catalogs. In the latter case, we extracted a synthetic source extension, in the same way as we extract their coordinates. {The source size is defined slightly differently in the two cases: for HGPS sources the $\sigma$ of the fitted 2D-Gaussian is provided, while for LHAASO sources, the radius that contains 39\% of the flux, $R_{39}$. The two quantities coincide when the source profile is a 2D-Gaussian. }

\begin{figure}
\centering
    \includegraphics[height=5 cm ,width=7 cm]{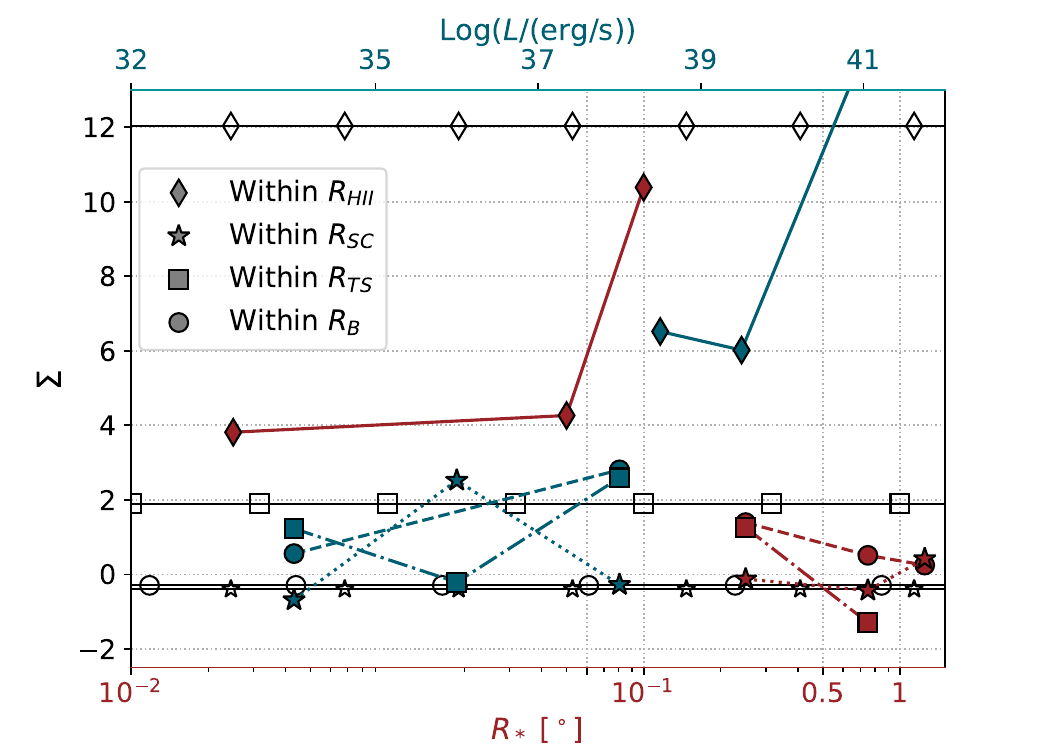}\includegraphics[height=5 cm ,width=7 cm]{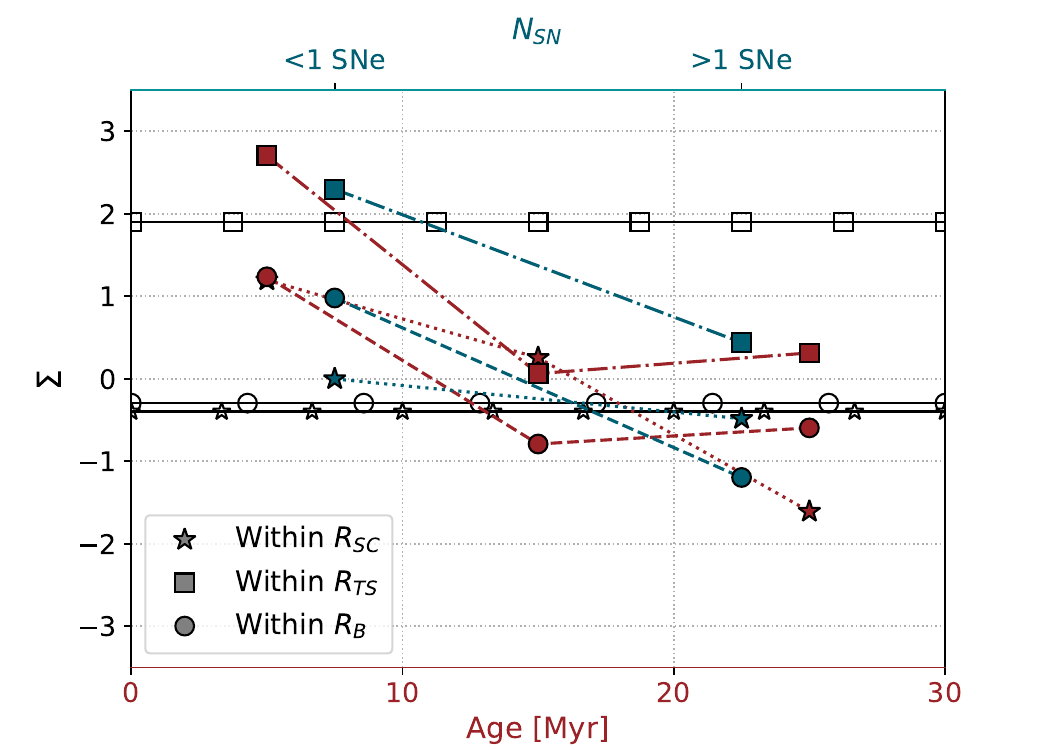} 
    \includegraphics[height=5 cm ,width=7 cm]{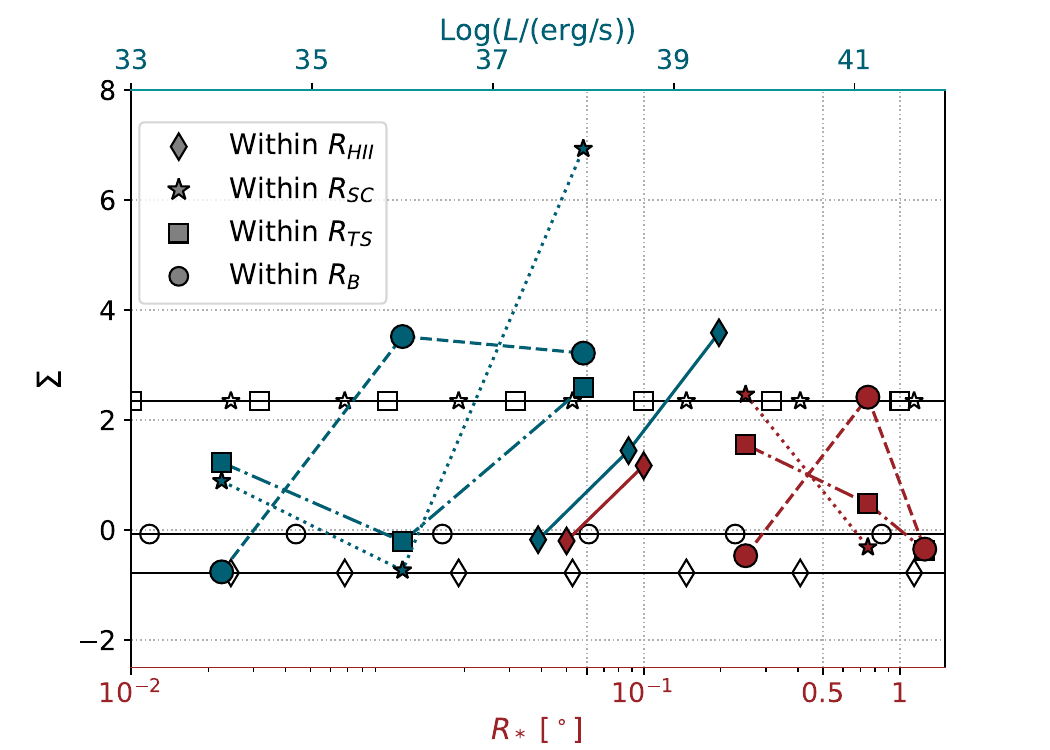}\includegraphics[height=5 cm ,width=7 cm]{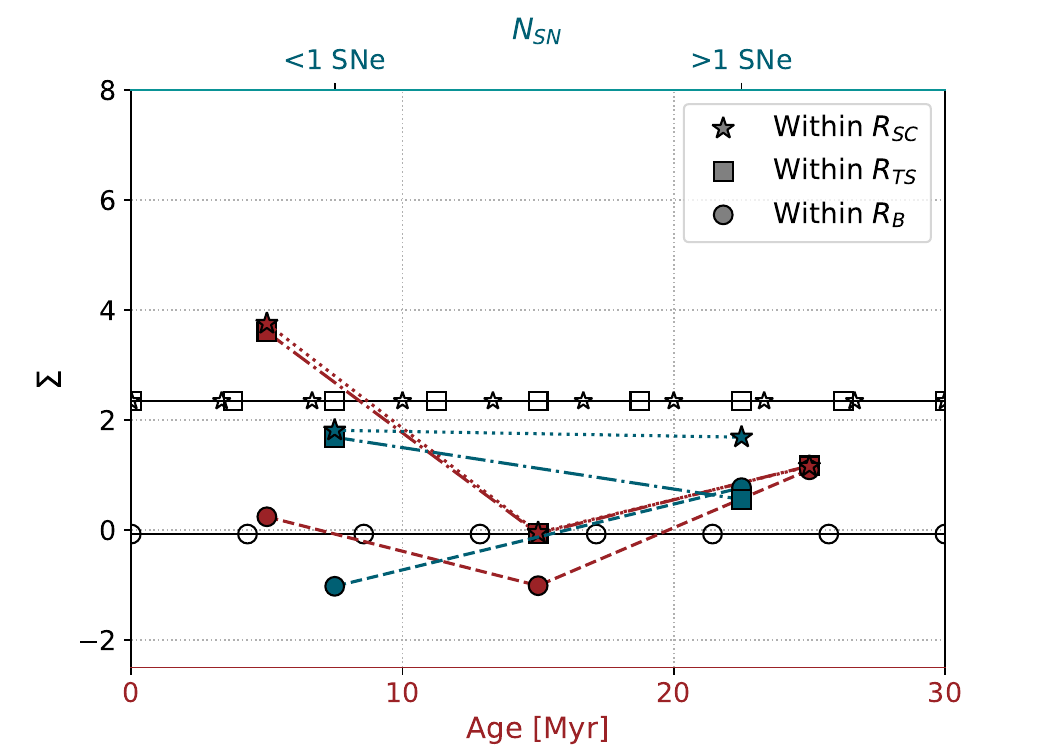}
    \includegraphics[height=5 cm ,width=7 cm]{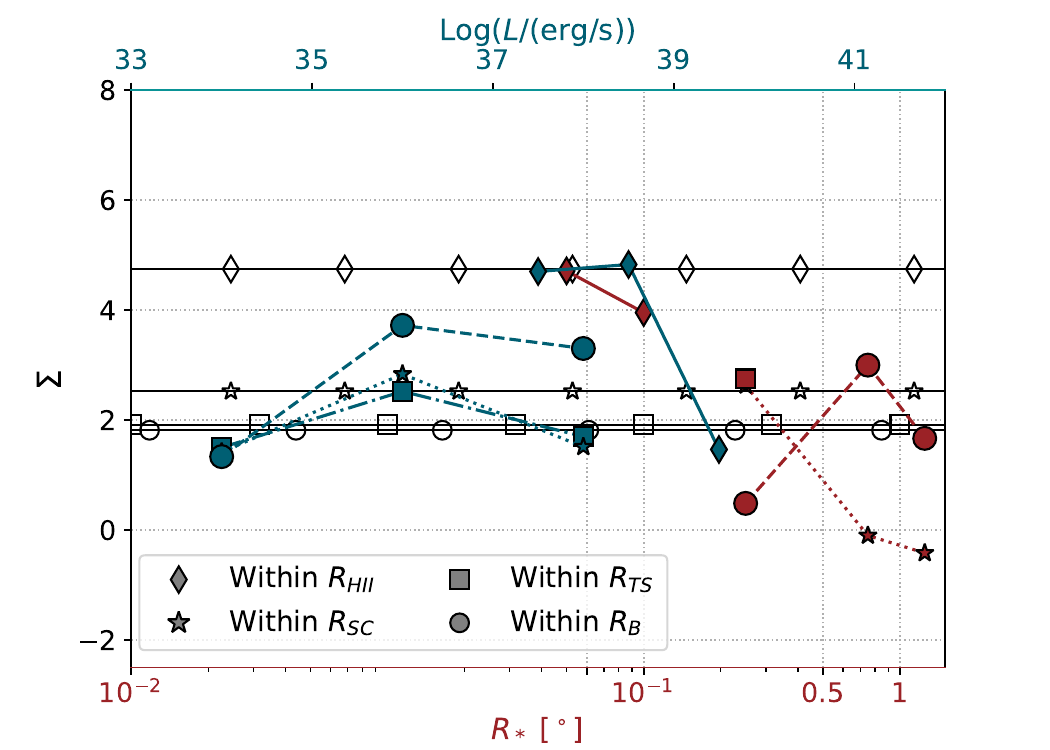}\includegraphics[height=5 cm ,width=7 cm]{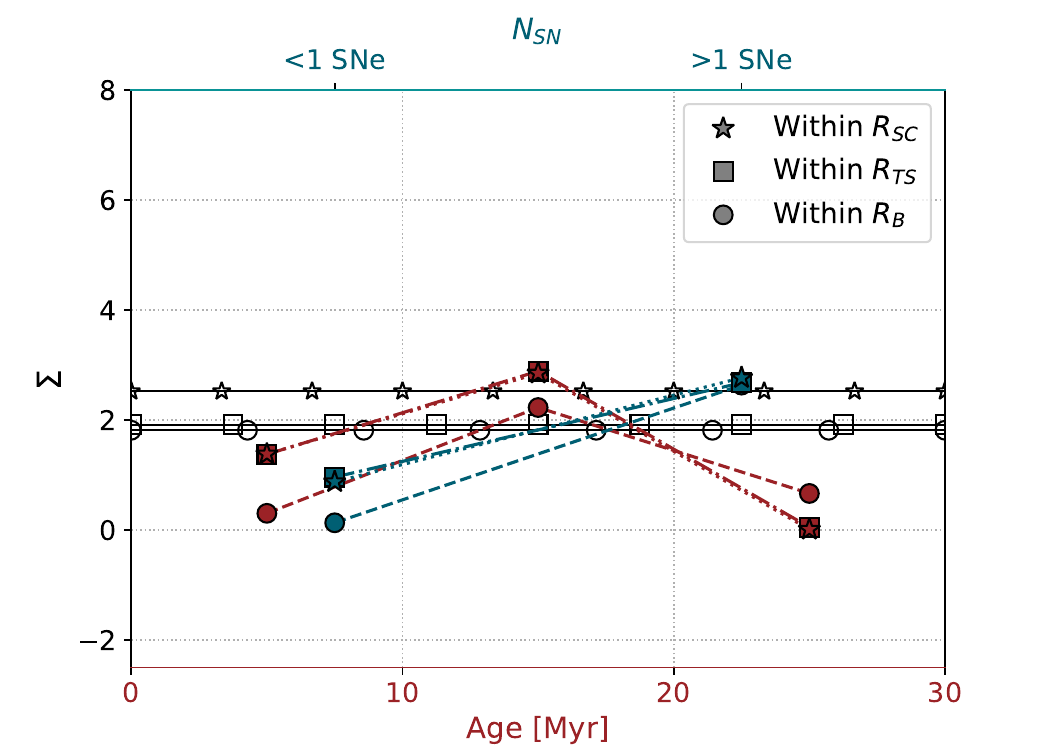}
    \includegraphics[height=5 cm ,width=7 cm]{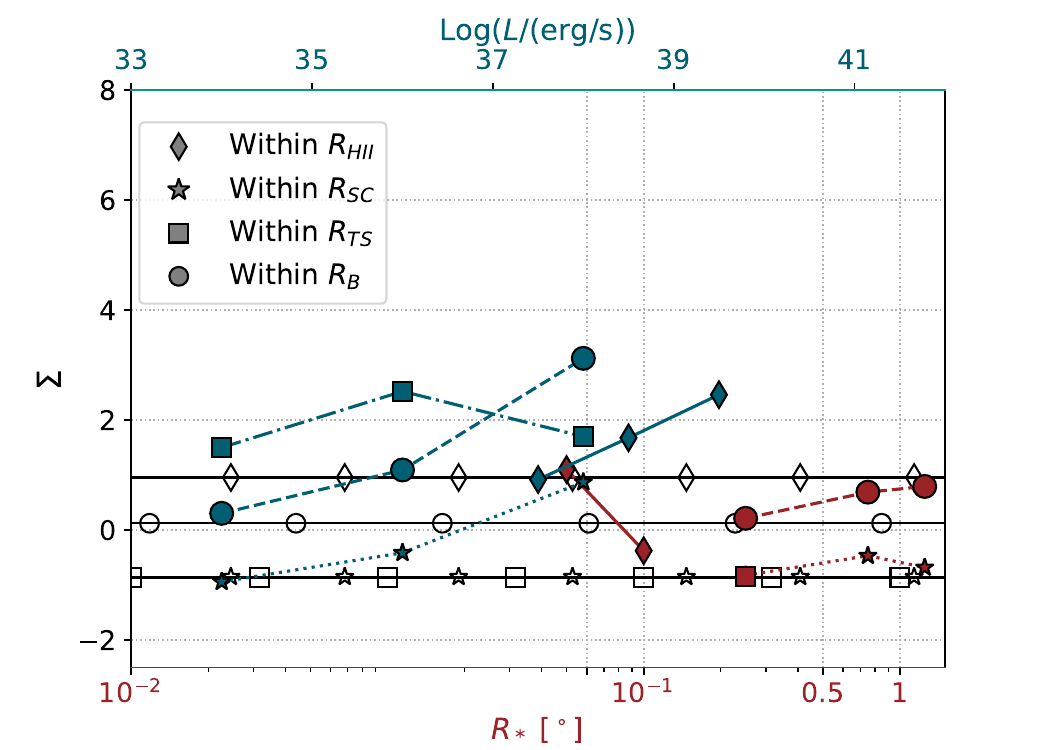}\includegraphics[height=5 cm ,width=7 cm]{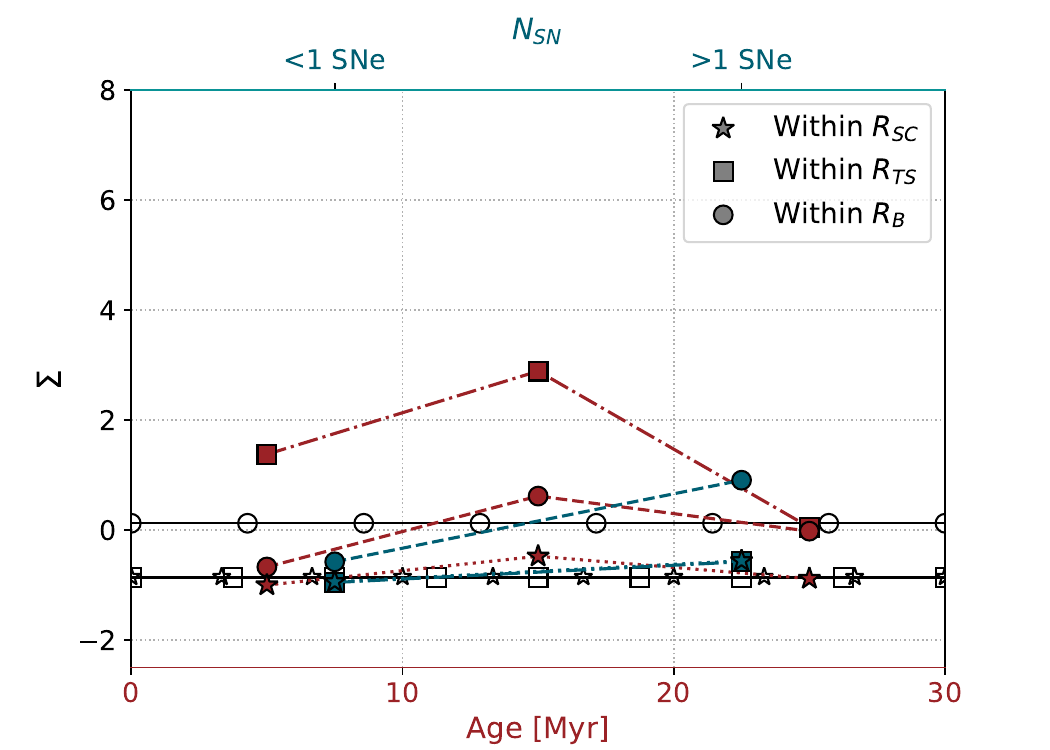}
    \caption{Significance of the correlation between SCs catalogs (horizontal lines) and their sub-samples divided by extension (red points) and luminosity (blue points), on the left, and by age (red points) and number of supernova explosions (blue points) on the right.  The indicated luminosity refers to the wind luminosity for Gaia SCs and to the IR luminosity for the WISE SCs.  Results obtained with WISE SCs are marked as diamond, while Gaia SCs are indicated by circles, squares and stars, depending on the matching radius considered among the bubble, the termination shock or the stellar radius, respectively.The different rows from top to the bottom report results obtained with the {\it Fermi}-LAT, H.E.S.S.-HGPS, LHAASO-WCDA, and LHAASO-KM2A respectively. Note that in the case of WISE SCs with $L_w>10^{39}$erg/s one match is found in the real catalog but zero matches are found in the 1000 realizations, therefore the matching significance is formally infinite, therefore it falls out of the plot axis. }
    \label{fig:sig_TeV}
\end{figure}

When dealing with the real catalog, we define the matching radius as the largest between the cluster and the gamma-ray source radius for each couple of objects compared; when matching the synthetic catalogs, we define it as the maximum between the cluster radius and the radius of the closest simulated cluster. 
The results of the matching significance are summarized in Fig. \ref{fig:sig_all} and Fig. \ref{fig:sig_TeV} for the different experiments and with the same sub-sample selection used for matching the \textit{Fermi}-LAT catalog. We can see that, as the size of the sources increases, the number of matching SCs increases, but alongside the matching-significance, $\Sigma$, decreases, suggesting that the associations are probably random (Fig. \ref{fig:sig_TeV}).
In general, with TeV sources the significance of association is rather small, exceeding a value of 2 only when considering the brightest ($L_w \gtrsim 10^{37}$ erg/s) SCs. A clear trend with age or number of SN events instead does not emerge. TeV sources are big objects that occupy large portions of the Galactic Plane, making the association much more complicated {as each of them intersects multiple SCs, especially when matching with the small WISE regions. We report the numbers of each case in the Appendix \ref{sec:nmatch}. 
In the case of HGPS sources, the association significance is further challenged by the selection in latitudes: indeed, in the region sampled by this survey, the density of WISE regions is almost ten times larger than in the rest of the Sky, plausibly causing a non-significant association between H.E.S.S. and WISE sources. Similar values are recorded when SCs are compared with KM2A sources. Differently, a significant ($\Sigma \sim 4$) correlation emerges with WCDA sources, although the results from the different sub-samples do not allow us to draw a firm conclusion on this point, as the significance decreases (even if slightly) both with the radius and with the luminosity. With Gaia sources instead the significance is always small $\lesssim 3$, with some exception towards very luminous SCs, which would need to be further investigated.    }

\section{Discussion and conclusions}
\begin{figure}
    \centering
    \includegraphics[width=0.7\linewidth]{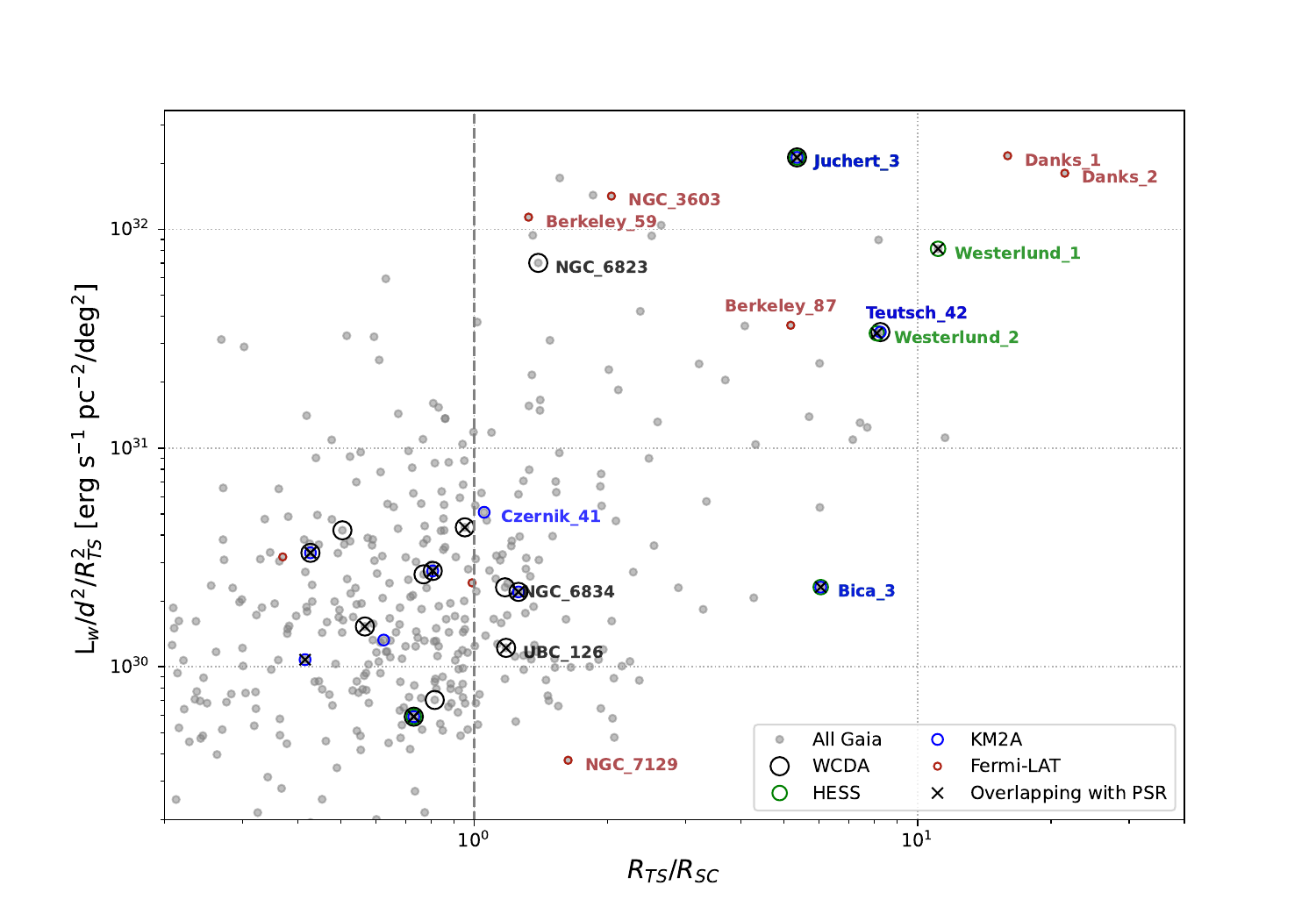}
    \caption{Surface brightness (in arbitrary units) against compacteness, expressed as the ratio between $R_{\rm TS}$ and $R_{\rm SC}$, for the Gaia SCs. The different colors indicate if the clusters overlap a source of a given catalog, as indicated in the legend. Sources that contain also a pulsar are indicated with a black cross. The  SC's name is indicated for matching compact ($R_{\rm TS}/R_{\rm SC} >1$) star clusters.}
    \label{fig:gaia_matched}
\end{figure}

 {The large extension $\gtrsim 1^{\circ}$  of Gaia SCs prevents us to statistically assess them as gamma-ray emitters. However, {some SCs in the Gaia sample may still emit gamma-rays} (see also \cite{Mitchell2024}). We tested which are the most probable associations, by considering the SC surface brightness, here assumed to be proportional to a fraction of $L_w$ and enclosed within an area of radius $R_{\rm TS}$, and its compactness ($R_{\rm TS}/R_{\rm SC}$). Indeed, to allow the collective termination shock to form, $R_{\rm TS}$ must exceed the radius of the stellar cluster core. The results are plotted in Fig. \ref{fig:gaia_matched}, where we indicate also those cases where a pulsar overlaps the considered source and hence could be responsible for the observed emission. {The brightest and less compact clusters (in the upper-right sector of the plot) seem to have the largest probability to be detected {and indeed most of them are found coincident with at least one gamma-ray source}. The undetected ones may lack of target material. However, all of these should be studied case by case, given also the uncertainty in the parameters estimates, especially concerning $R_{\rm TS}$.}
}

We report instead on a convincing correlation (Fig. \ref{fig:sig_all} and \ref{fig:sig_TeV}) between \textit{Fermi}-LAT sources and SCs in H\textsc{ii} regions,
probably favored by their locations in compact regions of dense gas. The large density that characterizes these regions also  {suggests a hadronic origin of the gamma-ray emission,}
even though a leptonic component cannot be excluded \textit{a priori}, given the large radiation fields characterizing the systems. {Indeed synchrotron emission, although at different frequencies, was also detected in coincidence with H\textsc{II} regions \citep{PadovaniHII,HIIXray} }.

Moreover, since H\textsc{ii} regions trace very young ($<3$~Myr) massive SCs, where no SN has occurred yet and, consequently, no pulsar is formed, the origin of the gamma-ray emission must be ascribed to the stars.
 
Interestingly, the large number of embedded SCs potentially associated to gamma-ray sources suggests that population studies may be more promising with these objects than in the case of SNRs, where only few tens of objects show clear gamma-ray emission \citep{Acero2016SNR, Abdollahi2022IncrementalCatalog}. At a first glance such a conclusion may be surprising, in that young SCs should be subdominant sources of CRs \citep{Seo2018TheProduction}. However, their larger lifetime as accelerators ($\sim$Myr vs. few $10^4$\,yr for SNRs) may favor their detection. Even when those regions are not resolved, they could contribute to the diffuse gamma-rays emission. Interestingly, \cite{Liu2022APlane} noted that the slope of the diffuse gamma-ray emission from the H\textsc{ii} component of gas is harder than the one in coincidence with the neutral gas. This fact may be explained as due to CRs freshly accelerated in embedded SCs, that have a spectrum harder than the diffuse CR component.

\newpage

{\it {\bf Acknowledgements.} We thank Dr. S. Celli and co-authors for sharing the values for Gaia SCs derived in their work, Dr. C. Amato for suggestions on the statistical analysis, and Dr. N. Bucciantini for helpful discussion. G.P. is supported by the INAF Astrophysical fellowship initiative. Part of this work was conducted in the context of AP's master thesis, partly financed by the Action pluriannuelle incitative-Astrophysique des processus de Hautes Énergies 2022-24 of the Observatoires de Paris, to which GP, SC and AP are grateful. {GM and EA are partially supported by INAF through Theory Grant 681 2022 “{\it Star Clusters As Cosmic Ray Factories}’’ and PRIN INAF 2019 “From massive stars to supernovae and supernova remnants: driving mass, energy and cosmic rays in our Galaxy’’. This work was partially supported by the European Union – NextGenerationEU RRF M4C2 1.1 under grant PRIN-MUR 2022TJW4EJ.}}

\bibliography{sample631}{}

\begin{thebibliography}{}
\expandafter\ifx\csname natexlab\endcsname\relax\def\natexlab#1{#1}\fi
\providecommand{\url}[1]{\href{#1}{#1}}
\providecommand{\dodoi}[1]{doi:~\href{http://doi.org/#1}{\nolinkurl{#1}}}
\providecommand{\doeprint}[1]{\href{http://ascl.net/#1}{\nolinkurl{http://ascl.net/#1}}}
\providecommand{\doarXiv}[1]{\href{https://arxiv.org/abs/#1}{\nolinkurl{https://arxiv.org/abs/#1}}}

\bibitem[{Abdalla {et~al.}(2018)Abdalla, Abramowski, Aharonian, Ait~Benkhali, Ang{\"{u}}ner, Arakawa, Arrieta, Aubert, Backes, Balzer, Barnard, Becherini, Becker~Tjus, Berge, Bernhard, Bernl{\"{o}}hr, Blackwell, B{\"{o}}ttcher, Boisson, Bolmont, Bonnefoy, Bordas, Bregeon, Brun, Brun, Bryan, B{\"{u}}chele, Bulik, Capasso, Carrigan, Caroff, Carosi, Casanova, Cerruti, Chakraborty, Chaves, Chen, Chevalier, Colafrancesco, Condon, Conrad, Davids, Decock, Deil, Devin, De~Wilt, Dirson, Djannati-Ata{\"{i}}, Domainko, Donath, Drury, Dutson, Dyks, Edwards, Egberts, Eger, Emery, Ernenwein, Eschbach, Farnier, Fegan, Fernandes, Fiasson, Fontaine, F{\"{o}}rster, Funk, F{\"{u}}{\ss}ling, Gabici, Gallant, Garrigoux, Gast, Gat{\'{e}}, Giavitto, Giebels, Glawion, Glicenstein, Gottschall, Grondin, Hahn, Haupt, Hawkes, Heinzelmann, Henri, Hermann, Hinton, Hofmann, Hoischen, Holch, Holler, Horns, Ivascenko, Iwasaki, Jacholkowska, Jamrozy, Jankowsky, Jankowsky, Jingo, Jouvin, Jung-Richardt, Kastendieck, Katarzy{\'{n}}ski,
  Katsuragawa, Katz, Kerszberg, Khangulyan, Kh{\'{e}}lifi, King, Klepser, Klochkov, Kl{\'{u}}zniak, Komin, Kosack, Krakau, Kraus, Kr{\"{u}}ger, Laffon, Lamanna, Lau, Lees, Lefaucheur, Lemi{\`{e}}re, Lemoine-Goumard, Lenain, Leser, Lohse, Lorentz, Liu, L{\'{o}}pez-Coto, Lypova, Marandon, Malyshev, Marcowith, Mariaud, Marx, Maurin, Maxted, Mayer, Meintjes, Meyer, Mitchell, Moderski, Mohamed, Mohrmann, Mor{\aa}, Moulin, Murach, Nakashima, De~Naurois, Ndiyavala, Niederwanger, Niemiec, Oakes, O'Brien, Odaka, Ohm, Ostrowski, Oya, Padovani, Panter, Parsons, Paz~Arribas, Pekeur, Pelletier, Perennes, Petrucci, Peyaud, Piel, Pita, Poireau, Poon, Prokhorov, Prokoph, P{\"{u}}hlhofer, Punch, Quirrenbach, Raab, Rauth, Reimer, Reimer, Renaud, De~Los~Reyes, Rieger, Rinchiuso, Romoli, Rowell, Rudak, Rulten, Safi-Harb, Sahakian, Saito, Sanchez, Santangelo, Sasaki, Schandri, Schlickeiser, Sch{\"{u}}ssler, Schulz, Schwanke, Schwemmer, Seglar-Arroyo, Settimo, Seyffert, Shafi, Shilon, Shiningayamwe, Simoni, Sol, Spanier,
  Spir-Jacob, {Stawarz}, Steenkamp, Stegmann, Steppa, Sushch, Takahashi, Tavernet, Tavernier, Taylor, Terrier, Tibaldo, Tiziani, Tluczykont, Trichard, Tsirou, Tsuji, Tuffs, Uchiyama, Van Der~Walt, Van~Eldik, Van~Rensburg, Van~Soelen, Vasileiadis, Veh, Venter, Viana, Vincent, Vink, Voisin, V{\"{o}}lk, Vuillaume, Wadiasingh, Wagner, Wagner, Wagner, White, Wierzcholska, Willmann, W{\"{o}}rnlein, Wouters, Yang, Zaborov, Zacharias, Zanin, Zdziarski, Zech, Zefi, Ziegler, Zorn, \& Zywucka}]{Abdalla2018}
Abdalla, H., Abramowski, A., Aharonian, F., {et~al.} 2018, Astronomy and Astrophysics, 612, A1, \dodoi{10.1051/0004-6361/201732098}

\bibitem[{{Abdo} {et~al.}(2010){Abdo}, {Ackermann}, {Ajello}, {Baldini}, {Ballet}, {Barbiellini}, {Bastieri}, {Bellazzini}, {Blandford}, {Bloom}, {Bonamente}, {Borgland}, {Bouvier}, {Brandt}, {Bregeon}, {Brigida}, {Bruel}, {Buehler}, {Buson}, {Caliandro}, {Cameron}, {Caraveo}, {Carrigan}, {Casandjian}, {Charles}, {Chaty}, {Chekhtman}, {Cheung}, {Chiang}, {Ciprini}, {Claus}, {Cohen-Tanugi}, {Conrad}, {Decesar}, {Dermer}, {de Palma}, {Digel}, {Silva}, {Drell}, {Dubois}, {Dumora}, {Favuzzi}, {Fortin}, {Frailis}, {Fukazawa}, {Fusco}, {Gargano}, {Gasparrini}, {Gehrels}, {Germani}, {Giglietto}, {Giordano}, {Glanzman}, {Godfrey}, {Grenier}, {Grondin}, {Grove}, {Guillemot}, {Guiriec}, {Hadasch}, {Harding}, {Hays}, {Jean}, {J{\'o}hannesson}, {Johnson}, {Johnson}, {Kamae}, {Katagiri}, {Kataoka}, {Kerr}, {Kn{\"o}dlseder}, {Kuss}, {Lande}, {Latronico}, {Lee}, {Lemoine-Goumard}, {Llena Garde}, {Longo}, {Loparco}, {Lovellette}, {Lubrano}, {Makeev}, {Mazziotta}, {Michelson}, {Mitthumsiri}, {Mizuno}, {Monte}, {Monzani},
  {Morselli}, {Moskalenko}, {Murgia}, {Naumann-Godo}, {Nolan}, {Norris}, {Nuss}, {Ohsugi}, {Omodei}, {Orlando}, {Ormes}, {Pancrazi}, {Parent}, {Pepe}, {Pesce-Rollins}, {Piron}, {Porter}, {Rain{\`o}}, {Rando}, {Reimer}, {Reimer}, {Reposeur}, {Ripken}, {Romani}, {Roth}, {Sadrozinski}, {Saz Parkinson}, {Sgr{\`o}}, {Siskind}, {Smith}, {Spinelli}, {Strickman}, {Suson}, {Takahashi}, {Takahashi}, {Tanaka}, {Thayer}, {Thayer}, {Tibaldo}, {Torres}, {Tosti}, {Tramacere}, {Uchiyama}, {Usher}, {Vasileiou}, {Venter}, {Vilchez}, {Vitale}, {Waite}, {Wang}, {Webb}, {Winer}, {Yang}, {Ylinen}, {Ziegler}, \& {Fermi LAT Collaboration}}]{GlobularSCs}
{Abdo}, A.~A., {Ackermann}, M., {Ajello}, M., {et~al.} 2010, \aap, 524, A75, \dodoi{10.1051/0004-6361/201014458}

\bibitem[{Abdollahi {et~al.}(2022)Abdollahi, Acero, Baldini, Ballet, Bastieri, Bellazzini, Berenji, Berretta, Bissaldi, Blandford, Bloom, Bonino, Brill, Britto, Bruel, Burnett, Buson, Cameron, Caputo, Caraveo, Castro, Chaty, Cheung, Chiaro, Cibrario, Ciprini, Coronado-Bl{\'{a}}zquez, Crnogorcevic, Cutini, D’Ammando, De~Gaetano, Digel, Di~Lalla, Dirirsa, Di~Venere, Dom{\'{i}}nguez, Fallah~Ramazani, Fegan, Ferrara, Fiori, Fleischhack, Franckowiak, Fukazawa, Funk, Fusco, Galanti, Gammaldi, Gargano, Garrappa, Gasparrini, Giacchino, Giglietto, Giordano, Giroletti, Glanzman, Green, Grenier, Grondin, Guillemot, Guiriec, Gustafsson, Harding, Hays, Hewitt, Horan, Hou, J{\'{o}}hannesson, Karwin, Kayanoki, Kerr, Kuss, Landriu, Larsson, Latronico, Lemoine-Goumard, Li, Liodakis, Longo, Loparco, Lott, Lubrano, Maldera, Malyshev, Manfreda, Mart{\'{i}}-Devesa, Mazziotta, Mereu, Meyer, Michelson, Mirabal, Mitthumsiri, Mizuno, Moiseev, Monzani, Morselli, Moskalenko, Negro, Nuss, Omodei, Orienti, Orlando, Paneque, Pei,
  Perkins, Persic, Pesce-Rollins, Petrosian, Pillera, Poon, Porter, Principe, Rain{\`{o}}, Rando, Rani, Razzano, Razzaque, Reimer, Reimer, Reposeur, S{\'{a}}nchez-Conde, Saz~Parkinson, Scotton, Serini, Sgr{\`{o}}, Siskind, Smith, Spandre, Spinelli, Sueoka, Suson, Tajima, Tak, Thayer, Thompson, Torres, Troja, Valverde, Wood, \& Zaharijas}]{Abdollahi2022IncrementalCatalog}
Abdollahi, S., Acero, F., Baldini, L., {et~al.} 2022, The Astrophysical Journal Supplement Series, 260, \dodoi{10.3847/1538-4365/ac6751}

\bibitem[{{Acero} {et~al.}(2016){Acero}, {Ackermann}, {Ajello}, {Baldini}, {Ballet}, {Barbiellini}, {Bastieri}, {Bellazzini}, {Bissaldi}, {Blandford}, {Bloom}, {Bonino}, {Bottacini}, {Brandt}, {Bregeon}, {Bruel}, {Buehler}, {Buson}, {Caliandro}, {Cameron}, {Caputo}, {Caragiulo}, {Caraveo}, {Casandjian}, {Cavazzuti}, {Cecchi}, {Chekhtman}, {Chiang}, {Chiaro}, {Ciprini}, {Claus}, {Cohen}, {Cohen-Tanugi}, {Cominsky}, {Condon}, {Conrad}, {Cutini}, {D'Ammando}, {de Angelis}, {de Palma}, {Desiante}, {Digel}, {Di Venere}, {Drell}, {Drlica-Wagner}, {Favuzzi}, {Ferrara}, {Franckowiak}, {Fukazawa}, {Funk}, {Fusco}, {Gargano}, {Gasparrini}, {Giglietto}, {Giommi}, {Giordano}, {Giroletti}, {Glanzman}, {Godfrey}, {Gomez-Vargas}, {Grenier}, {Grondin}, {Guillemot}, {Guiriec}, {Gustafsson}, {Hadasch}, {Harding}, {Hayashida}, {Hays}, {Hewitt}, {Hill}, {Horan}, {Hou}, {Iafrate}, {Jogler}, {J{\'o}hannesson}, {Johnson}, {Kamae}, {Katagiri}, {Kataoka}, {Katsuta}, {Kerr}, {Kn{\"o}dlseder}, {Kocevski}, {Kuss}, {Laffon}, {Lande},
  {Larsson}, {Latronico}, {Lemoine-Goumard}, {Li}, {Li}, {Longo}, {Loparco}, {Lovellette}, {Lubrano}, {Magill}, {Maldera}, {Marelli}, {Mayer}, {Mazziotta}, {Michelson}, {Mitthumsiri}, {Mizuno}, {Moiseev}, {Monzani}, {Moretti}, {Morselli}, {Moskalenko}, {Murgia}, {Nemmen}, {Nuss}, {Ohsugi}, {Omodei}, {Orienti}, {Orlando}, {Ormes}, {Paneque}, {Perkins}, {Pesce-Rollins}, {Petrosian}, {Piron}, {Pivato}, {Porter}, {Rain{\`o}}, {Rando}, {Razzano}, {Razzaque}, {Reimer}, {Reimer}, {Renaud}, {Reposeur}, {Rousseau}, {Saz Parkinson}, {Schmid}, {Schulz}, {Sgr{\`o}}, {Siskind}, {Spada}, {Spandre}, {Spinelli}, {Strong}, {Suson}, {Tajima}, {Takahashi}, {Tanaka}, {Thayer}, {Thompson}, {Tibaldo}, {Tibolla}, {Torres}, {Tosti}, {Troja}, {Uchiyama}, {Vianello}, {Wells}, {Wood}, {Wood}, {Yassine}, {den Hartog}, \& {Zimmer}}]{Acero2016SNR}
{Acero}, F., {Ackermann}, M., {Ajello}, M., {et~al.} 2016, \apjs, 224, 8, \dodoi{10.3847/0067-0049/224/1/8}

\bibitem[{Anderson {et~al.}(2014)Anderson, Bania, Balser, Cunningham, Wenger, Johnstone, \& Armentrout}]{Anderson2014TheRegions}
Anderson, L.~D., Bania, T.~M., Balser, D.~S., {et~al.} 2014, Astrophysical Journal, Supplement Series, 212, 1, \dodoi{10.1088/0067-0049/212/1/1}

\bibitem[{Buzzoni(2002)}]{Buzzoni2002UltravioletSynthesis}
Buzzoni, A. 2002, The Astronomical Journal, 123, \dodoi{10.1086/338896}

\bibitem[{{Bykov} {et~al.}(2020){Bykov}, {Marcowith}, {Amato}, {Kalyashova}, {Kruijssen}, \& {Waxman}}]{Bykov2020}
{Bykov}, A.~M., {Marcowith}, A., {Amato}, E., {et~al.} 2020, \ssr, 216, 42, \dodoi{10.1007/s11214-020-00663-0}

\bibitem[{{Cantat-Gaudin} {et~al.}(2020){Cantat-Gaudin}, {Anders}, {Castro-Ginard}, {Jordi}, {Romero-G{\'o}mez}, {Soubiran}, {Casamiquela}, {Tarricq}, {Moitinho}, {Vallenari}, {Bragaglia}, {Krone-Martins}, \& {Kounkel}}]{Cantat-Gaudin2020}
{Cantat-Gaudin}, T., {Anders}, F., {Castro-Ginard}, A., {et~al.} 2020, \aap, 640, A1, \dodoi{10.1051/0004-6361/202038192}

\bibitem[{{Cao} {et~al.}(2024){Cao}, {Aharonian}, {An}, {Axikegu}, {Bai}, {Bao}, {Bastieri}, {Bi}, {Bi}, {Cai}, {Cao}, {Cao}, {Cao}, {Chang}, {Chang}, {Chen}, {Chen}, {Chen}, {Chen}, {Chen}, {Chen}, {Chen}, {Chen}, {Chen}, {Chen}, {Chen}, {Chen}, {Cheng}, {Cheng}, {Cui}, {Cui}, {Cui}, {Cui}, {Dai}, {Dai}, {Dai}, {Danzengluobu}, {Della Volpe}, {Dong}, {Duan}, {Fan}, {Fan}, {Fang}, {Fang}, {Feng}, {Feng}, {Feng}, {Feng}, {Feng}, {Gabici}, {Gao}, {Gao}, {Gao}, {Gao}, {Gao}, {Gao}, {Ge}, {Geng}, {Giacinti}, {Gong}, {Gou}, {Gu}, {Guo}, {Guo}, {Guo}, {Guo}, {Han}, {He}, {He}, {He}, {He}, {He}, {Heller}, {Hor}, {Hou}, {Hou}, {Hou}, {Hu}, {Hu}, {Hu}, {Huang}, {Huang}, {Huang}, {Huang}, {Huang}, {Huang}, {Huang}, {Ji}, {Jia}, {Jia}, {Jiang}, {Jiang}, {Jiang}, {Jin}, {Kang}, {Ke}, {Kuleshov}, {Kurinov}, {Li}, {Li}, {Li}, {Li}, {Li}, {Li}, {Li}, {Li}, {Li}, {Li}, {Li}, {Li}, {Li}, {Li}, {Li}, {Li}, {Li}, {Li}, {Li}, {Liang}, {Liang}, {Lin}, {Liu}, {Liu}, {Liu}, {Liu}, {Liu}, {Liu}, {Liu}, {Liu}, {Liu}, {Liu}, {Liu},
  {Liu}, {Liu}, {Liu}, {Lu}, {Luo}, {Lv}, {Ma}, {Ma}, {Ma}, {Mao}, {Min}, {Mitthumsiri}, {Mu}, {Nan}, {Neronov}, {Ou}, {Pang}, {Pattarakijwanich}, {Pei}, {Qi}, {Qi}, {Qiao}, {Qin}, {Ruffolo}, {S{\'a}iz}, {Semikoz}, {Shao}, {Shao}, {Shchegolev}, {Sheng}, {Shu}, {Song}, {Stenkin}, {Stepanov}, {Su}, {Sun}, {Sun}, {Sun}, {Tam}, {Tang}, {Tang}, {Tian}, {Wang}, {Wang}, {Wang}, {Wang}, {Wang}, {Wang}, {Wang}, {Wang}, {Wang}, {Wang}, {Wang}, {Wang}, {Wang}, {Wang}, {Wang}, {Wang}, {Wang}, {Wang}, {Wang}, {Wang}, {Wang}, {Wei}, {Wei}, {Wei}, {Wen}, {Wu}, {Wu}, {Wu}, {Wu}, {Wu}, {Xi}, {Xia}, {Xia}, {Xiang}, {Xiao}, {Xiao}, {Xin}, {Xin}, {Xing}, {Xiong}, {Xu}, {Xu}, {Xu}, {Xu}, {Xue}, {Yan}, {Yan}, {Yan}, {Yang}, {Yang}, {Yang}, {Yang}, {Yang}, {Yang}, {Yang}, {Yang}, {Yang}, {Yao}, {Yao}, {Ye}, {Yin}, {Yin}, {You}, {You}, {Yu}, {Yuan}, {Yue}, {Zeng}, {Zeng}, {Zeng}, {Zha}, {Zhang}, {Zhang}, {Zhang}, {Zhang}, {Zhang}, {Zhang}, {Zhang}, {Zhang}, {Zhang}, {Zhang}, {Zhang}, {Zhang}, {Zhang}, {Zhang}, {Zhang}, {Zhang},
  {Zhang}, {Zhang}, {Zhao}, {Zhao}, {Zhao}, {Zhao}, {Zhao}, {Zheng}, {Zhou}, {Zhou}, {Zhou}, {Zhou}, {Zhou}, {Zhou}, {Zhou}, {Zhu}, {Zhu}, {Zhu}, {Zhu}, {Zuo}, \& {(The Lhaaso Collaboration)}}]{Cao2023Catalog}
{Cao}, Z., {Aharonian}, F., {An}, Q., {et~al.} 2024, \apjs, 271, 25, \dodoi{10.3847/1538-4365/acfd29}

\bibitem[{{Celli} {et~al.}(2023){Celli}, {Specovius}, {Menchiari}, {Mitchell}, \& {Morlino}}]{Celli2023}
{Celli}, S., {Specovius}, A., {Menchiari}, S., {Mitchell}, A., \& {Morlino}, G. 2023, arXiv e-prints, arXiv:2311.09089, \dodoi{10.48550/arXiv.2311.09089}

\bibitem[{Cesarsky \& Montmerle(1983)}]{Cesarsky1983a}
Cesarsky, C.~J., \& Montmerle, T. 1983, Space Science Reviews, 36, 173, \dodoi{10.1007/BF00167503}

\bibitem[{{de O{\~n}a Wilhelmi} {et~al.}(2022){de O{\~n}a Wilhelmi}, {L{\'o}pez-Coto}, {Amato}, \& {Aharonian}}]{Emma2022}
{de O{\~n}a Wilhelmi}, E., {L{\'o}pez-Coto}, R., {Amato}, E., \& {Aharonian}, F. 2022, \apjl, 930, L2, \dodoi{10.3847/2041-8213/ac66cf}

\bibitem[{{Gabici}(2023)}]{Gabici2023SCs}
{Gabici}, S. 2023, arXiv e-prints, arXiv:2301.06505, \dodoi{10.48550/arXiv.2301.06505}

\bibitem[{Liu \& Yang(2022)}]{Liu2022APlane}
Liu, B., \& Yang, R.~Z. 2022, Astronomy and Astrophysics, 659, \dodoi{10.1051/0004-6361/202039759}

\bibitem[{Makai {et~al.}(2017)Makai, Anderson, Mascoop, \& Johnstone}]{Makai2017TheRegions}
Makai, Z., Anderson, L.~D., Mascoop, J.~L., \& Johnstone, B. 2017, The Astrophysical Journal, 846, 64, \dodoi{10.3847/1538-4357/aa84b6}

\bibitem[{{Menchiari}(2023)}]{Menchiari2023}
{Menchiari}, S. 2023, arXiv e-prints, arXiv:2307.03477, \dodoi{10.48550/arXiv.2307.03477}

\bibitem[{{Mitchell} {et~al.}(2024){Mitchell}, {Morlino}, {Celli}, {Menchiari}, \& {Specovius}}]{Mitchell2024}
{Mitchell}, A. M.~W., {Morlino}, G., {Celli}, S., {Menchiari}, S., \& {Specovius}, A. 2024, arXiv e-prints, arXiv:2403.16650, \dodoi{10.48550/arXiv.2403.16650}

\bibitem[{Miville-Desch{\^{e}}nes {et~al.}(2016)Miville-Desch{\^{e}}nes, Murray, \& Lee}]{Miville-Deschenes2016}
Miville-Desch{\^{e}}nes, M.-A., Murray, N., \& Lee, E.~J. 2016, The Astrophysical Journal, 834, 57, \dodoi{10.3847/1538-4357/834/1/57}

\bibitem[{{Montmerle}(2010)}]{Montmerle2010}
{Montmerle}, T. 2010, in Astronomical Society of the Pacific Conference Series, Vol. 422, High Energy Phenomena in Massive Stars, ed. J.~{Mart{\'\i}}, P.~L. {Luque-Escamilla}, \& J.~A. {Combi}, 85, \dodoi{10.48550/arXiv.0909.0222}

\bibitem[{{Padovani} {et~al.}(2019){Padovani}, {Marcowith}, {S{\'a}nchez-Monge}, {Meng}, \& {Schilke}}]{PadovaniHII}
{Padovani}, M., {Marcowith}, A., {S{\'a}nchez-Monge}, {\'A}., {Meng}, F., \& {Schilke}, P. 2019, \aap, 630, A72, \dodoi{10.1051/0004-6361/201935919}

\bibitem[{Peron {et~al.}(2024)Peron, Casanova, Gabici, Baghmanyan, \& Aharonian}]{Peron2024ThePopulation}
Peron, G., Casanova, S., Gabici, S., Baghmanyan, V., \& Aharonian, F. 2024, Nature Astronomy, \dodoi{10.1038/s41550-023-02168-6}

\bibitem[{Saha {et~al.}(2020)Saha, Dom{\'{i}}nguez, Tibaldo, Marchesi, Ajello, Lemoine-Goumard, \& L{\'{o}}pez}]{Saha2020Morphological3603}
Saha, L., Dom{\'{i}}nguez, A., Tibaldo, L., {et~al.} 2020, The Astrophysical Journal, 897, \dodoi{10.3847/1538-4357/ab9ac2}

\bibitem[{Seo {et~al.}(2018)Seo, Kang, \& Ryu}]{Seo2018TheProduction}
Seo, J., Kang, H., \& Ryu, D. 2018, Journal of the Korean Astronomical Society, 51, 37, \dodoi{10.5303/JKAS.2018.51.2.37}

\bibitem[{{Stahler} \& {Palla}(2004)}]{Stahler2004}
{Stahler}, S.~W., \& {Palla}, F. 2004, {The Formation of Stars} (Wiley)

\bibitem[{Tatischeff {et~al.}(2021)Tatischeff, Raymond, Duprat, Gabici, \& Recchia}]{Tatischeff2021TheComposition}
Tatischeff, V., Raymond, J.~C., Duprat, J., Gabici, S., \& Recchia, S. 2021, Monthly Notices of the Royal Astronomical Society, 508, 1321, \dodoi{10.1093/MNRAS/STAB2533}

\bibitem[{{Townsley} {et~al.}(2011){Townsley}, {Broos}, {Chu}, {Gruendl}, {Oey}, \& {Pittard}}]{HIIXray}
{Townsley}, L.~K., {Broos}, P.~S., {Chu}, Y.-H., {et~al.} 2011, \apjs, 194, 16, \dodoi{10.1088/0067-0049/194/1/16}

\bibitem[{Vieu {et~al.}(2022)Vieu, Gabici, Tatischeff, \& Ravikularaman}]{Vieu2022CosmicSuperbubbles}
Vieu, T., Gabici, S., Tatischeff, V., \& Ravikularaman, S. 2022, Monthly Notices of the Royal Astronomical Society, 512, 1275, \dodoi{10.1093/mnras/stac543}

\bibitem[{Vieu \& Reville(2023)}]{Vieu2023MassiveEnergies}
Vieu, T., \& Reville, B. 2023, Monthly Notices of the Royal Astronomical Society, 519, \dodoi{10.1093/mnras/stac3469}

\bibitem[{Weaver {et~al.}(1977)Weaver, McCray, Castor, Shapiro, Moore, Weaver, McCray, Castor, Shapiro, \& Moore}]{Weaver1977InterstellarEvolution.}
Weaver, R., McCray, R., Castor, J., {et~al.} 1977, ApJ, 218, 377, \dodoi{10.1086/155692}

\bibitem[{{Yang} \& {Aharonian}(2017)}]{yang2016NGC}
{Yang}, R.-z., \& {Aharonian}, F. 2017, \aap, 600, A107, \dodoi{10.1051/0004-6361/201630213}

\bibitem[{{Yang} \& {Wang}(2020)}]{Yang2020}
{Yang}, R.-Z., \& {Wang}, Y. 2020, \aap, 640, A60, \dodoi{10.1051/0004-6361/202037518}

\end{thebibliography}
\bibliographystyle{aasjournal}

\appendix

\section{Monte Carlo extractions}\label{sec:montecarlo}
\begin{figure}
    \centering
    \includegraphics[width=1 \linewidth]{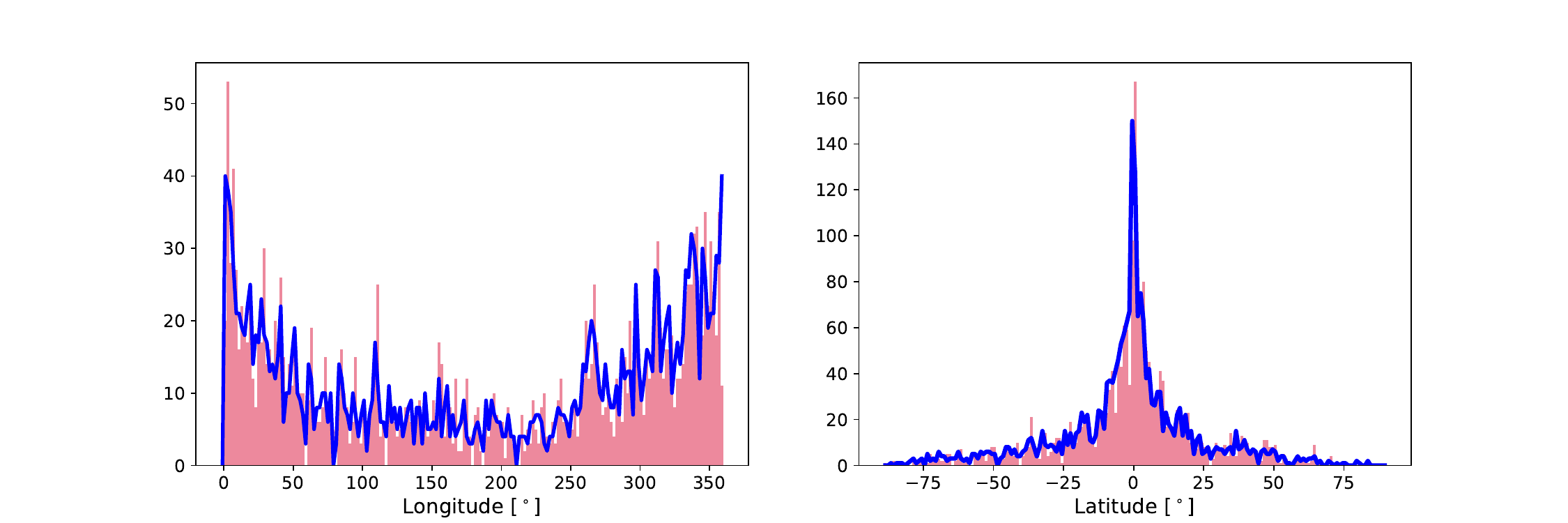}
    \caption{Longitude and latitude distribution of the {\it Fermi}-LAT 2082 non associated sources (blue lines) compared to the same distributions of one realization of synthetic clusters (red bars).}
    \label{fig:dist}
\end{figure}

\begin{figure}
    \centering
    \includegraphics[width=1 \linewidth]{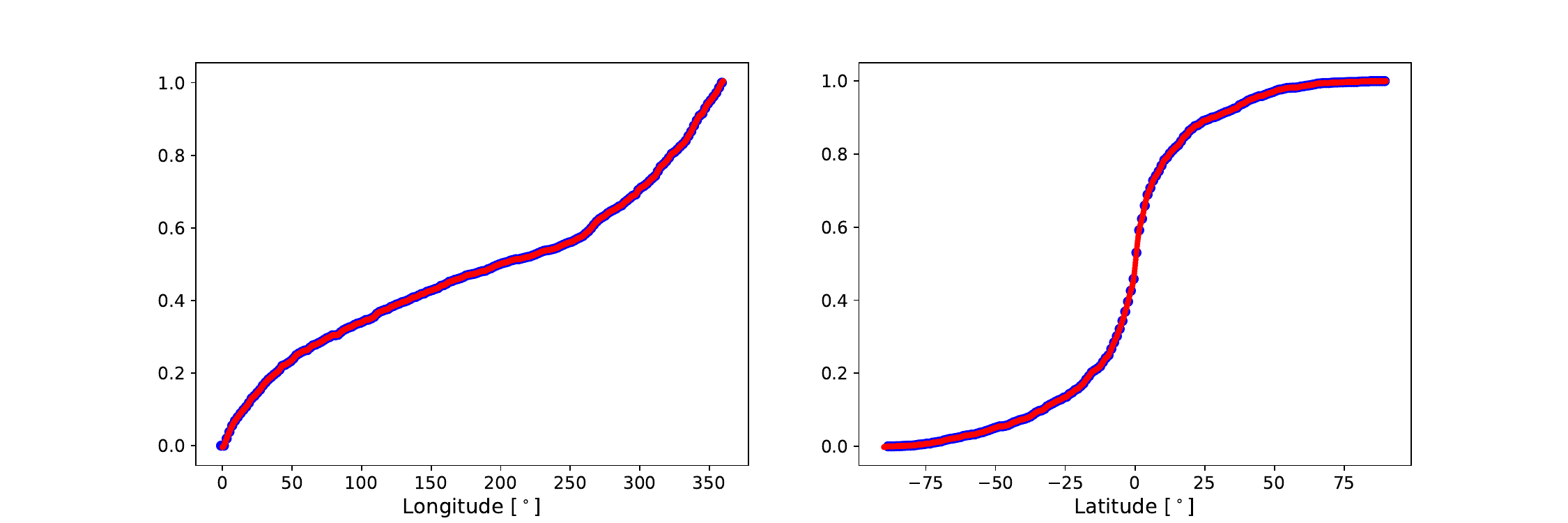}
    \caption{Longitude and latitude cumulative distribution of the {\it Fermi}-LAT 2082 non associated sources.}
    \label{fig:cum_dist}
\end{figure}

For each considered gamma-ray catalog, we generated 1000 synthetic catalogs with the same number of objects, by extracting over the distribution $\mathcal{F}(x)$ of the interested variable $x$, that in our case are the longitude, $l$, and latitude, $b$, for {\it Fermi}-LAT sources, and extensions, $R_\gamma$, for TeV sources. We proceeded then in the following way. First we sample the distributions with steps of 2$^\circ$, 1$^\circ$ and 0.01$^\circ$ respectively for $l$,$b$, and $R_{\gamma}$. As an example, we show in Fig. \ref{fig:dist} the distribution of $l$ and $b$ of the {\it Fermi}-LAT unidentified sources. From that, we derive the cumulative distribution, $\mathcal{C}(x)$ , for each variable as:
\begin{equation}  \label{eq:appA1}
  \mathcal{C}(x) = \int^x_{x_{\min}} \mathcal{F}(x') dx' \,, 
\end{equation}
where $x_{\min}$ is the smallest values of the considered variable.
The cumulative distribution is then normalized between 0 and 1 and is interpolated to a finer (0.01$^\circ$) grid (Figure \ref{fig:cum_dist}). To derive the new variables needed to compose the synthetic catalogs, we perform a random extraction over a uniform distribution of values comprised between 0 and 1. Each of the extracted points, $C_{ext}$, corresponds to a value of the normalized cumulative distribution, that in turn corresponds to a unique value for $x$. If a value of $C_{ext}$ falls between two sampled values of the distributions, namely within $\mathcal{C}(x_1)$ and $\mathcal{C}(x_2)$,  instead of choosing the closest $x$ that would give the value of $\mathcal{C}(x)=C_{ext}$, we approximate the distribution as the line that connects the two points in order to avoid discontinuities and choose $x$ that gives the corresponding value on the connecting line. 

\begin{figure}
    \centering
    \includegraphics[width=0.8 \linewidth]{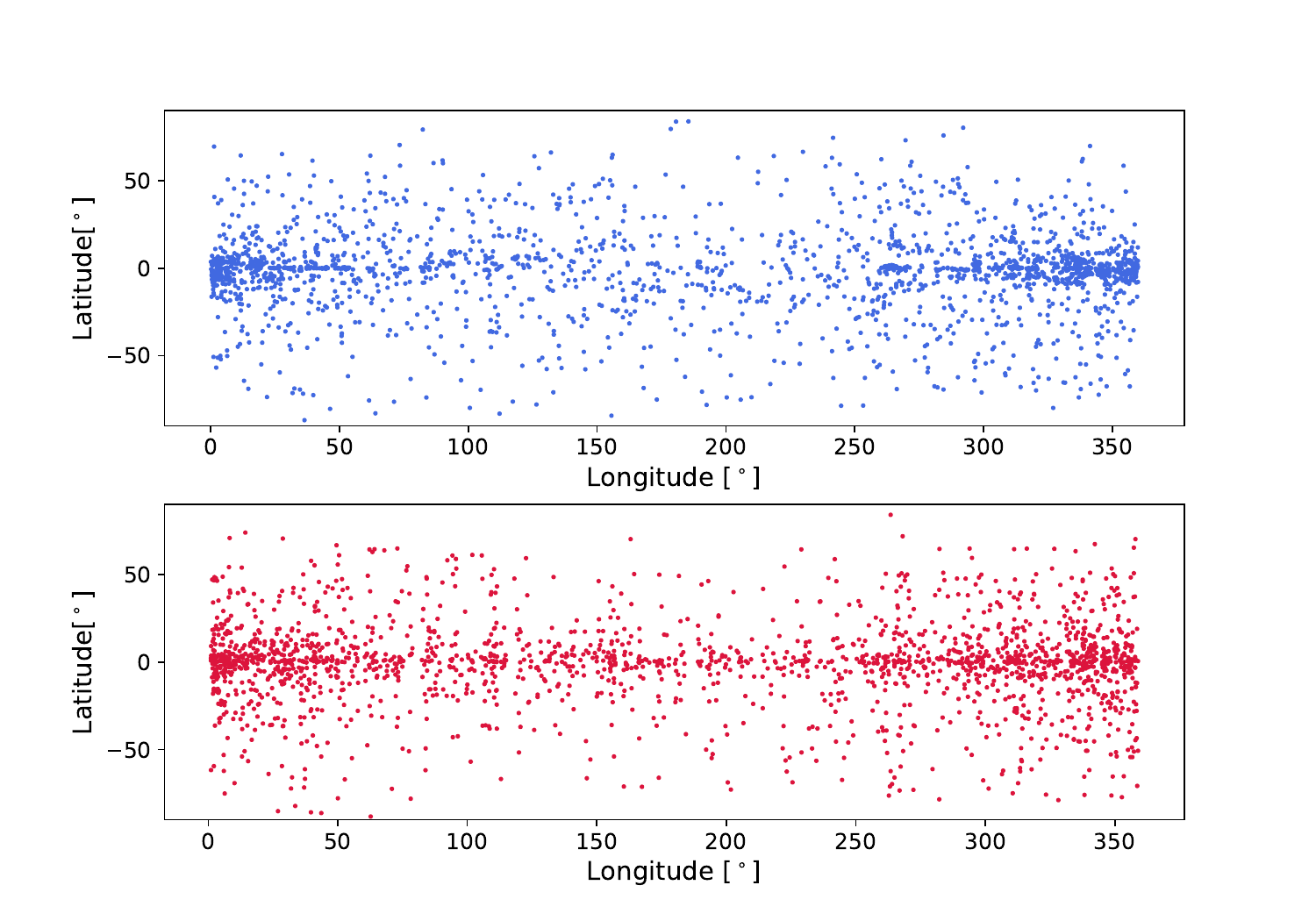}
    \caption{Longitude and latitude distribution of the real {\it Fermi}-LAT sources (blue) and of one simulated catalog (red).}
    \label{fig:sky_dist}
\end{figure}

The results is a new localization of sources that follow the same distribution as the starting one, as can be seen from the red bars in Fig. \ref{fig:dist} and from the comparison of the maps of real and simulated catalog in Fig. \ref{fig:sky_dist}. 


\subsection{Chance-correlation}
\begin{figure}
    \centering
    \includegraphics[width=0.45\linewidth]{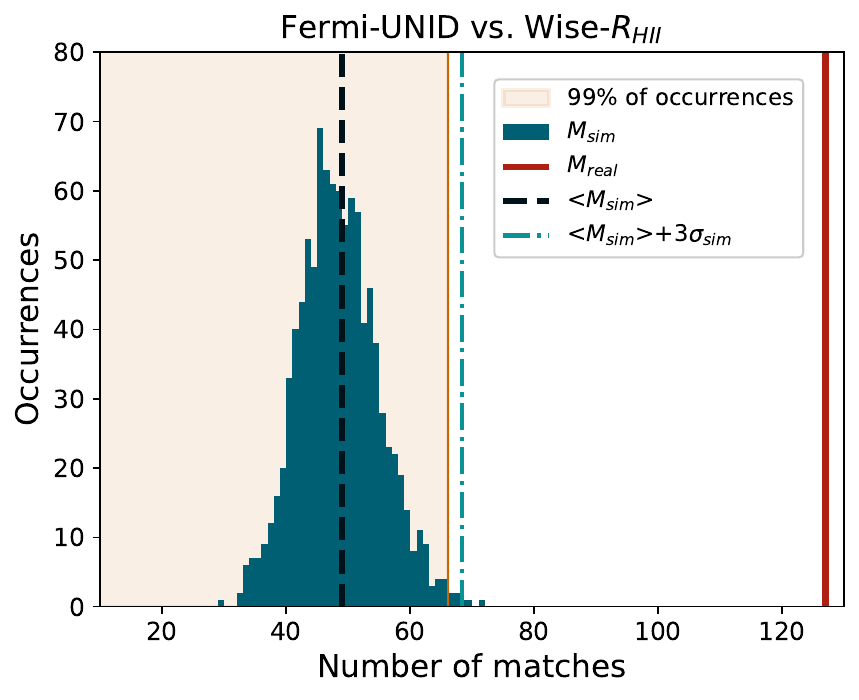}\includegraphics[width=0.45\linewidth]{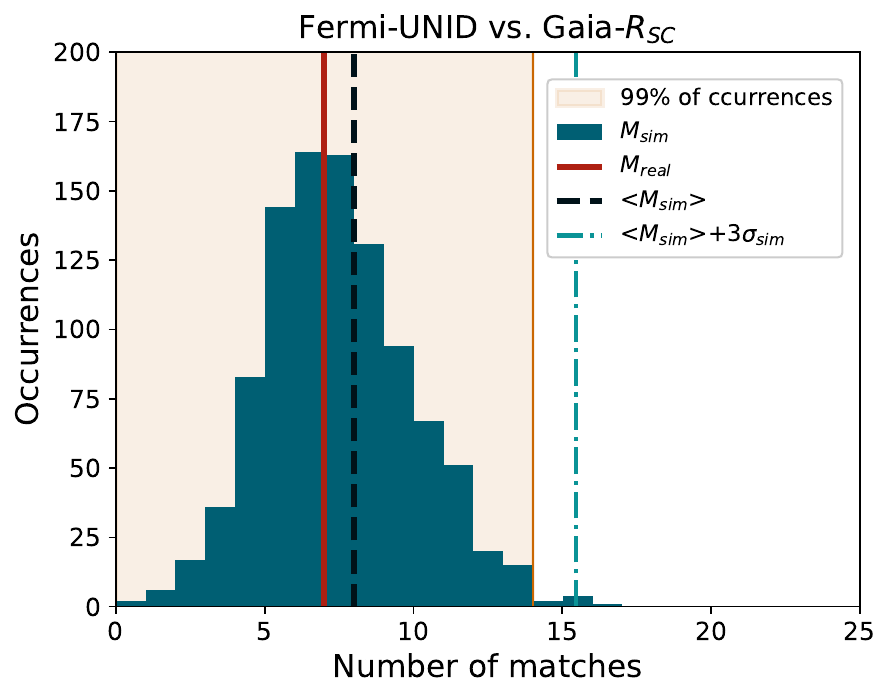}
    \includegraphics[width=0.45\linewidth]{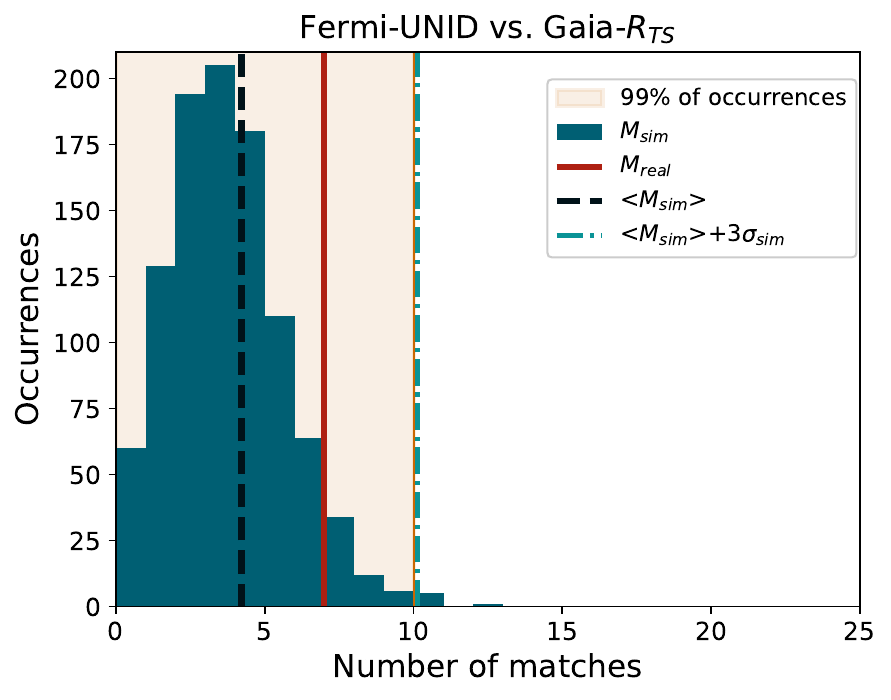}\includegraphics[width=0.45\linewidth]{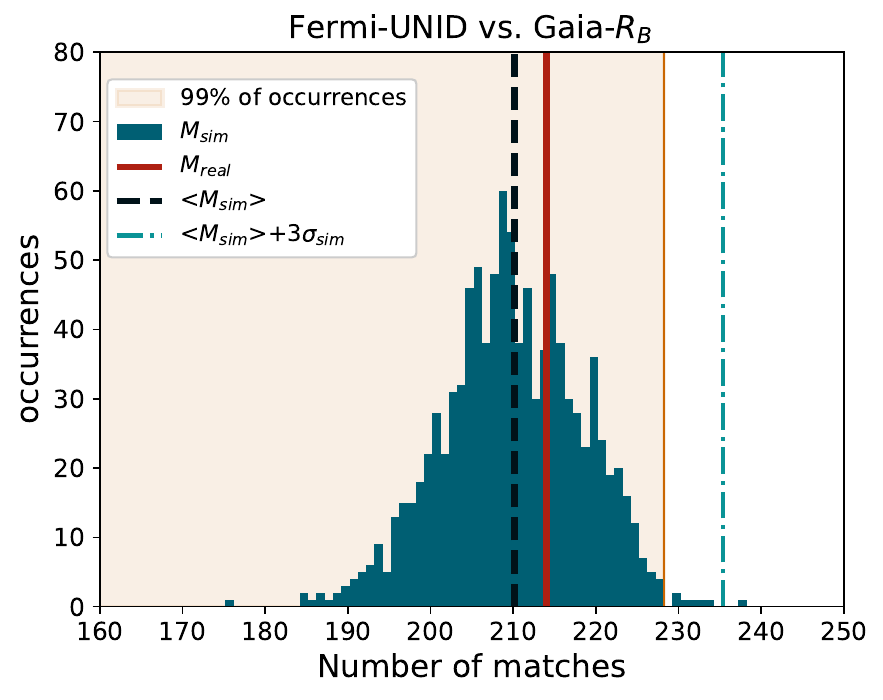}
    \caption{{Distribution of matches, $M_{sim}$, of the simulated catalogs, with the different SC catalogs, and within different matching radii, as indicated in each panel. The black dashed line indicates the average of $M_{sim}$; the yellow area indicates the zone where 99\% of occurrences are contained; the dot-dashed light-blue line indicates the events at 3 standard deviations ($\sigma_{sim}$) from the average; the red line indicates the number of real matches.   }}
    \label{fig:hist}
\end{figure}

The synthetic catalogs are used then to determine the degree of chance correlation in the associations. 
To test the validity of this method we repeated the matching procedure with real and synthetic catalogs of different source classes from the ones discussed in the main text. In particular, we matched the \textit{Fermi}-LAT identified blazars and the considered SC (WISE and Gaia) catalogs, since the two source classes are  expected to be completely uncorrelated. The resulting significance with 1000 simulations with the WISE and Gaia catalog is always $<-1$, except when using the bubble radius as the matching radius for the Gaia catalog, which is slightly larger $\sim-0.4$, likely due to the large extension of the bubbles. 
Instead, to test two source classes that are expected to correlate, we tested the \textit{Fermi}-LAT identified pulsars against the H.E.S.S. identified pulsar wind nebulae, using their extension as matching radius and we obtained a significance of $\gtrsim 8$. These values can be considered a benchmark when comparing the results obtained in our investigation. {For the case of \textit{Fermi}-LAT, we show in Fig. \ref{fig:hist}, the distribution of the number of matches obtained with the simulated, $M_{sim}$, and real catalogs, $M_{real}$ for the different considered cluster catalogs and matching radii. We indicate in the figure the threshold within which one can find 99\% of the occurrences. We also show the average, and the 3-sigma deviation from it. The latter is always somewhat above the 99\% threshold, like in the Gaussian case where the 3-sigma level represents the 99.7\% of occurrences. This can be used to compare the resulting significance of the different samples evaluated in the main text. In general we would define correlated, two catalogs that have $\Sigma>3$, as the probability of chance coincidence in that case is smaller than 1\%. }

\section{Properties of the analyzed samples}
\begin{figure}
    \centering
    \includegraphics[width=0.5 \linewidth]{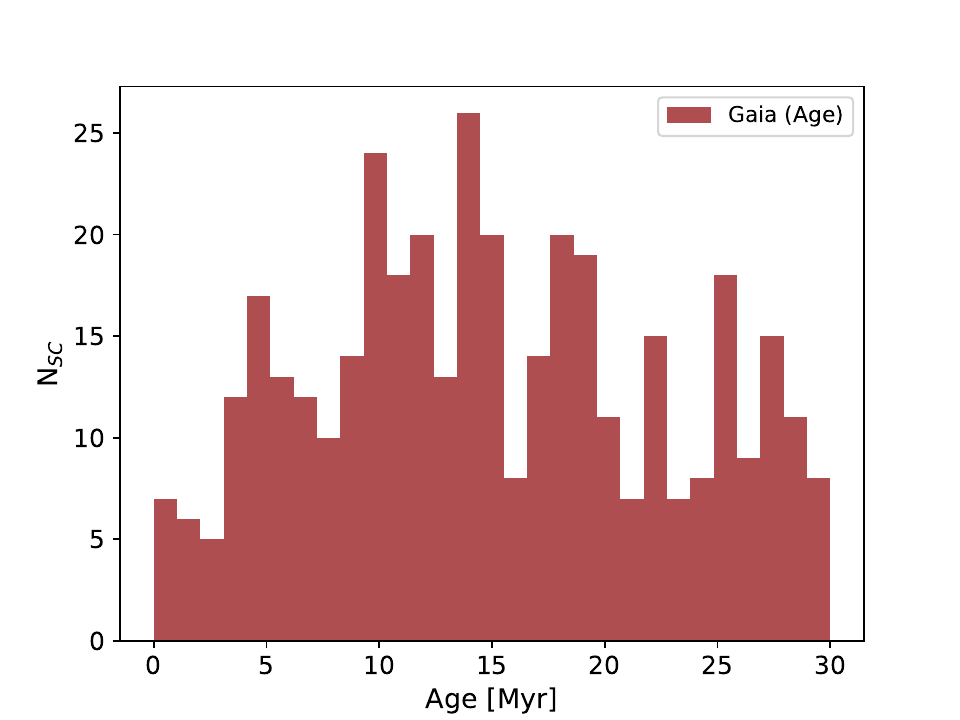}\includegraphics[width=0.5 \linewidth]{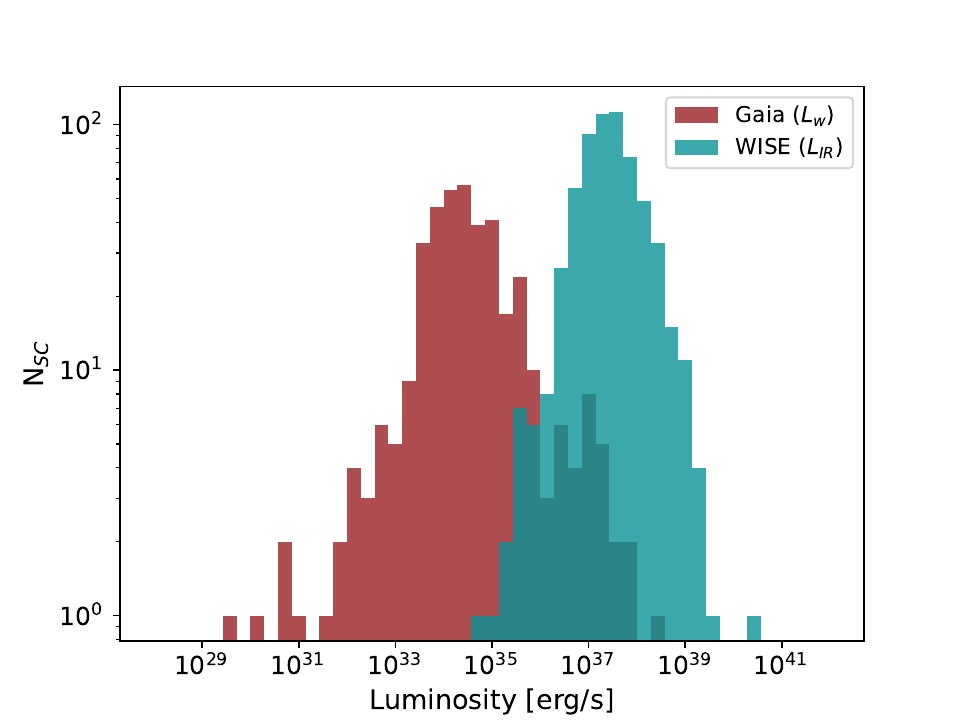}
    \caption{Physical properties of the analyzed sample of SCs that are considered in the discussion: \textit{on the left:} the age distribution of Gaia sample; \textit{on the right:} the distribution of wind luminosity as calculated by \cite{Celli2023}, and of IR luminosity of WISE H\textsc{ii} regions, for the available objects. }
    \label{fig:phys_stats}
\end{figure}

 {In the following (see Fig. \ref{fig:phys_stats}), we show the distribution of the physical properties of the clusters considered in this work. For Gaia SCs we know the age and wind luminosity as derived by \cite{Celli2023}, for WISE regions instead we have the infrared luminosity for a sample of SCs that we use as a proxy of the stellar power. }

\subsection{Supernova explosion rate in SCs} \label{sec:supernovae}
To compute the probability of a supernova (SN) event in a SC, we follow the prescription by \cite{Celli2023} that we summarize here. This approach implies that a star undergoes a SN explosion when the cluster age exceeds the lifetime of a star of that mass. The lifetime of a star as a function of its mass follows the empirical relation  derived for example in \cite{Buzzoni2002UltravioletSynthesis}: 
$$\log_{10}{(M_{max}/M_{\odot})} = -C \ln[\log_{10}(T_{age}/yr)/A-B]$$

where A = 4.35, B = 1.30, C = 1.101.

We assume that the mass distribution in the clusters follows the average mass distribution in the Galaxy, the so-called Kroupa-relation: 
\begin{equation} \label{eq:Kroupa_dis}
\frac{dN}{dM}\propto
\begin{cases}
    M^{-\alpha_1} & \mathrm{if}~ 0.08\,M_\odot<M<0.5\,M_\odot \\
    M^{-\alpha_2} &  \mathrm{if}~ 0.5\,M_\odot<M<1\,M_\odot \\
    M^{-\alpha_3} &  \mathrm{if}~ M>1\,M_\odot \\
\end{cases}    
\end{equation}
with $\alpha_1=0.3$, $\alpha_2=1.3$, and $\alpha_3=2.3$. We further put a constraint on the maximum mass allowed for a star in a SC of a given mass as done in \cite{Celli2023}. This prevents to overpredict the number of SNe in those low-mass SCs, where very massive stars are not expected at all.

\begin{figure}
    \centering
    \includegraphics[width=0.5\linewidth]{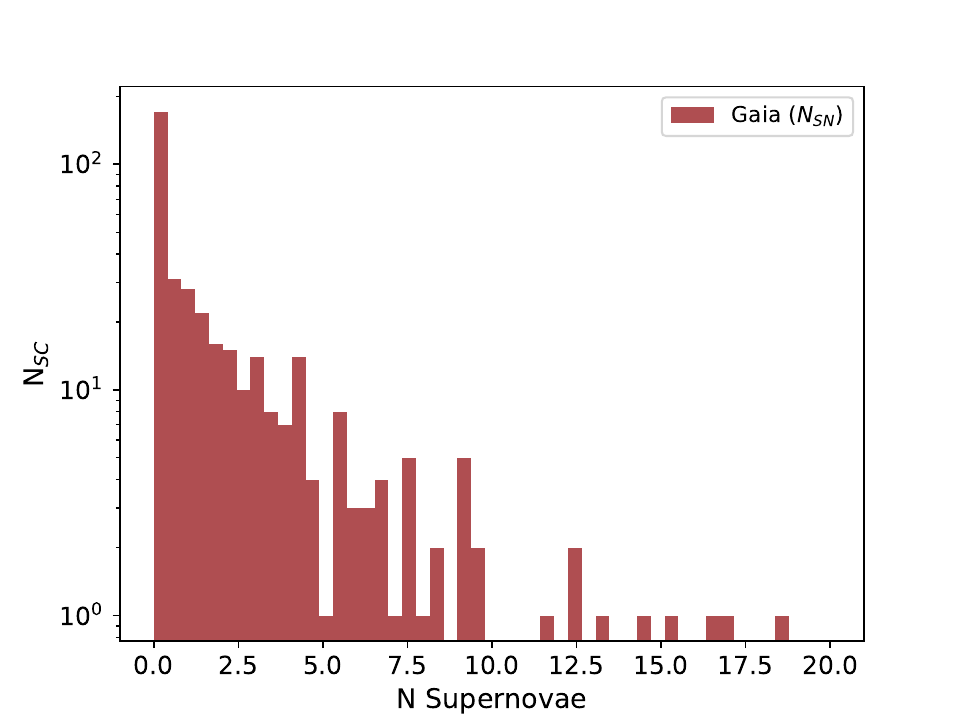}
    \caption{Distribution of the Gaia SCs in terms of number of supernova explosions estimated to have occured during the age of the SCs}
    \label{fig:n_sn}
\end{figure}

We then calculate the fraction of stars for which the expected lifetime is shorter than the age of the cluster.  {The resulting distribution of number of SNe believed to have occurred during the entire age of the clusters, is plotted in Figure~\ref{fig:n_sn}}

\section{Number of matching}\label{sec:nmatch}
 {In Table \ref{tab:nSCs} we report the number of matching sources. As most of the times multiple sources overlap, we report both the number of SCs from the different catalogs having at least one overlapping gamma-ray sources within the tested matching radius, and the number of gamma-ray sources overlapping at least one SC with the same criteria. We also indicate the number of gamma-ray sources that are free from GeV pulsars. To check that we cross-matched the list of interesting sources with the list of the GeV identified pulsars contained in the 4FGL catalog \citep{Abdollahi2022IncrementalCatalog}, but we cannot exclude that other (non-detected in GeV) pulsars are present.    }

\begin{table}[]
    \centering
    
    \begin{tabular}{l|cccc|cccc}
        Catalog & N$_{\rm SC}$-$R_{\rm HII}$ &  N$_{\rm SC}$-$R_{\rm SC}$ &  N$_{\rm SC}$-$R_{\rm TS}$ &  N$_{\rm SC}$-$R_{\rm B}$ & N$_{\gamma}$-$R_{\rm HII}$&  N$_{\gamma}$-$R_{\rm SC}$&  N$_{\gamma}$-$R_{\rm TS}$&  N$_{\gamma}$-$R_{\rm B}$  \\
        \hline \hline

          {\it Fermi}-LAT & 127(125) & 7(7) & 8(8) & {214}(0)& 138(138) & 9(9) & 8(8)  & 666(666) \\
          H.E.S.S. &  307(305)&  5(5)& 5(4)& 33(6)&  27(21)&  5(1)& 5(1)& 34(24)\\
          LHAASO-WCDA &  959(955)&  14(14)& 14(14)& 41(18)&  42(23)&  12(5)& 12(5)& 33(17)\\
          LHAASO-KM2A &  832(827)&  10(10)& 10(10)& 50(21)&  49(29)&  9(2)& 9(2)& 41(22)\\
          
    \end{tabular}
    \caption{Number of matching sources: both the number of SCs, $N_{\rm SC}$, overlapping at least one gamma-ray source and the number of gamma-ray sources, $N_{\gamma}$, that overlap at least a SC within the different matching radii. We indicate in parentheses the number of sources that are free from pulsar contamination. This coincides with the total for Fermi sources as they are point-like; for TeV sources instead we considered their reported extension as searching radius for PSRs therefore decreasing the number of free sources.}
    \label{tab:nSCs}
\end{table}

\end{document}